\documentclass{sig-alternate}
\usepackage{fancyhdr} 
\usepackage{lastpage} 
\usepackage{extramarks}
\usepackage{graphics} 
\usepackage{graphicx}
\usepackage{color}
\usepackage{epsfig} 
\usepackage{mathptmx} 
\usepackage{times} 
\usepackage{amsmath} 
\usepackage{amssymb}  
\usepackage{float}
\usepackage{caption}
\usepackage{lipsum}
\usepackage{tikz}
\usepackage{bm}
\usepackage{array}
\usepackage{xhfill}
\definecolor{bluish_green_coop}{rgb}{0, 0.6, 0.5}
\definecolor{orange_defec}{rgb}{0.9019, 0.6235, 0}

\newcommand{\tikzcircle}[2][red,fill=red]{\tikz[baseline=-0.5ex]\draw[#1,radius=#2] (0,0) circle ;}%
\newcommand\crule[3][black]{\textcolor{#1}{\rule{#2}{#3}}}

\makeatletter
\newcommand{\thickhline}{%
    \noalign {\ifnum 0=`}\fi \hrule height 2pt
    \futurelet \reserved@a \@xhline
}
\newcolumntype{"}{@{\hskip\tabcolsep\vrule width 2pt\hskip\tabcolsep}}
\makeatother

\begin{document}



\newfont{\mycrnotice}{ptmr8t at 7pt}
\newfont{\myconfname}{ptmri8t at 7pt}
\let\crnotice\mycrnotice%
\let\confname\myconfname%

\permission{Copyright 2016 Association for Computing Machinery. ACM acknowledges that this contribution was authored or co-authored by an employee, contractor or affiliate of a national government. As such, the Government retains a nonexclusive, royalty-free right to publish or reproduce this article, or to allow others to do so, for Government purposes only.}
\conferenceinfo{BCB'16,}{October 02--05, 2016, Seattle, WA, USA.}
\copyrightetc{Copyright 2016 ACM. \the\acmcopyr}
\crdata{ISBN 978-1-4503-4225-4/16/10\ ...\$15.00.\\
DOI: http://dx.doi.org/10.1145/2975167.2975190}

\clubpenalty= 10000
\widowpenalty = 10000



%
\conferenceinfo{BCB '16}{October 2--5, 2016, Seattle, WA, USA}

\title{Stability Analysis of Population Dynamics Model in Microbial Biofilms with Non-participating Strains}
%
%
%
%
%

%
\author{
%
%
Jeet Banerjee$^{1}$, Tanvi Ranjan$^{2}$, Ritwik Kumar Layek$^{3}$\\
       \affaddr{Systems Biology and Computer Vision Laboratory, Dept. of Electronics \& Electrical Communication Engineering}\\
       \affaddr{Indian Institute of Technology Kharagpur, India 721302}\\
       \email{\{jeetbanerjee$^{1}$, ritwik$^{3}$\}@ece.iitkgp.ernet.in, tanviranjan@iitkgp.ac.in$^{2}$}
}

\maketitle
\begin{abstract}
The existence of phenotypic heterogeneity in single-species bacterial biofilms is well-established in the published literature. However, the modeling of population dynamics in biofilms from the viewpoint of social interactions, i.e. interplay between heterotypic strains, and the analysis of this kind using control theory are not addressed significantly. Therefore, in this paper, we theoretically analyze the population dynamics model in microbial biofilms with \textit{non-participating strains} (coexisting with public goods producers and non-producers) in the context of evolutionary game theory and nonlinear dynamics. Our analysis of the replicator dynamics model is twofold: first without the inclusion of spatial pattern, and second with the consideration of degree of assortment. In the first case, Lyapunov stability analysis of the stable equilibrium point of the proposed replicator system determines $(1,0)$ (`full dominance of cooperators') as a global asymptotic stable equilibrium whenever the return exceeds the metabolic cost of cooperation. Hence, the global asymptotic stable nature of $(1,0)$ in the context of non-consideration of spatial pattern helps to justify mathematically the adversity in the eradication of ``cooperative enterprise" that is an infectious biofilm. In the second case, we found non-existence of global asymptotic stability in the system, and it unveils two additional phenomena - bistability and coexistence. In this context, two inequality conditions are derived for the `full dominance of cooperators' and coexistence. Therefore, the inclusion of spatial pattern in biofilms with non-competing strains intends conditional dominance of pathogenic (with respect to the hosts) public goods producers which can be an effective strategy towards the control of an infectious biofilm with the drug-dependent regulation of degree of segregation. Furthermore, the simulation results of the proposed dynamics for both the discussed scenario confirm the results of the analysis of equilibrium points. The proposed stability analysis not only demonstrate a mathematical framework to analyze the population dynamics in biofilms but also gives a clue to control an infectious biofilm, where phenotypic and spatial heterogeneity exist. 
\end{abstract}
\category{J.3}{Life and Medical Sciences}{Biology and genetics, Health} \category{J.4}{Social and Behavioral Sciences}{Sociology} \category{I.6.4}{Model Validation and Analysis}{}   
\terms{Theory}
\keywords{Biofilm, Evolutionary game theory, Nonlinear dynamics, Population dynamics, Public goods, Social interactions, Spatial pattern}
\section{Introduction}
The previous idea of the unicellular behavior of microbes has been metamorphosed into the collective behavioral property \cite{shapiro1988bacteria, lopez2010biofilms}. In a natural habitat, bacteria commonly live in a surface-associated, matrix-enclosed (matrix of extracellular polymeric substances, EPS), complex differentiated communities, called biofilms \cite{lopez2010biofilms}. A plethora of cooperative processes, such as fruiting body formation \cite{shapiro1988bacteria}, summative secretion of EPS \cite{branda2005biofilms}, biosurfactants \cite{pamp2007multiple}, virulence factors \cite{vega2014collective}, etc. is required for a microbial community to remain viable. These collective phenomena are actively controlled and regulated by quorum sensing \cite{nadell2008evolution, heilmann2015bacteria} within a subpopulation of public goods producers, called cooperators \cite{celiker2013cellular}. However, phenotypic heterogeneity among clone-mates is often observed in biofilms \cite{czaran2009microbial}. This heterogeneity primarily controls the ability of constituent population to produce and use public goods. Cooperators produce public goods at a cost to themselves, whereas defective strains (defectors) use the produced goods without contribution to the shared pool \cite{hibbing2010bacterial}. It is naturally expected that defectors will thrive under evolutionary pressure due to the low cost of survival; however, the maintenance of cooperation in the colony is supported by the different mechanisms as discussed in the existing literature \cite{nowak2006five, xavier2011molecular, drescher2014solutions}. Spatial pattern formation is one of the existing mechanisms towards the persistence of cooperation in biofilms \cite{van2014density, ratzke2016self, martin2016laboratory}. Although, most of these hypotheses suggest that the synergistic and antagonistic interactions within and in-between of public goods producers and non-producers primarily regulates the stability of biofilms \cite{martin2016laboratory}.\\ 
\indent Population dynamics models depend on retention of public goods and metabolic cost of public goods production are discussed in \cite{hauert2006evolutionary, hauert2008ecological, wakano2009spatial} for ecological public goods games with constant size and only with cooperative and defective strains. However, a bacterial colony comprises other strains than cooperators and defectors \cite{czaran2009microbial}. Moreover, a majority of chronic infections \cite{bjarnsholt2013applying} are due to pathogenic biofilms \cite{li2012simulation, sharma2013simulating}. Therefore, the stability analysis of such biofilms is an important question to study towards the development of novel anti-biofilm drugs \cite{bjarnsholt2013applying, bjarnsholt2013role}. In this paper, we formulate the replicator dynamics model \cite{frey2010evolutionary, banerjee2015dynamics} in microbial biofilms with \textit{non-competing strains} \cite{czaran2009microbial} in two ways - first without consideration of spatial pattern and secondly include the pattern formation. Further, we analyze the stability of the equilibrium points of the proposed replicator system using nonlinear control theory \cite{strogatz2014nonlinear}.\\
\indent As a result of stability analysis of the proposed system in the absence of spatial pattern, we obtain the `full dominance of cooperators' (i.e. $(1,0)$) in the colony is only a globally asymptotically stable equilibrium whenever retention factor \cite{wakano2009spatial} is greater than the cost of cooperation, whereas, $(1,0)$ is a saddle-node whenever cost exceeds return. Hence, whenever retention exceeds metabolic cost, the cooperators take over the entire population. On the contrary, analysis of replicator equation with consideration of the level of assortment \cite{van2014density} concludes non-existence of global stability in the system, and it introduces bistability and coexistence. In this context, we derived the inequality conditions for the `sole dominance of producers' and coexistence. Therefore, the conditional dominance of cooperators in biofilms with \textit{non-competing strains} and spatial patterns gives a clue to control pathogenicity via regulation of degree of assortment. To the best of the author's knowledge, this paper presents the first of its kind stability analysis of a replicator system with and without the inclusion of the degree of segregation in a combined framework of evolutionary game theory (EGT) and nonlinear dynamics.\\             
\indent The organization of the paper is as follows. Section \ref{key_contri} discusses the contributions of our work. Section \ref{proposed_model} describes the proposed model, stability analysis of the replicator system without consideration of spatial pattern, and corresponding simulation results. The global asymptotic stability analysis of the equilibrium point $(1,0)$ using Lyapunov function candidate is also presented in this section. Section \ref{analysis_with_segre} presents the stability analysis of replicator equation with the inclusion of spatial pattern. In this section, detailed simulations of the dynamics for four different cases are also discussed. Section \ref{conclu} concludes the paper with a few inferences \& remarks on the proposed analysis and future scope.
\section{Key Contributions of the Work}
\label{key_contri}
$\bullet${ The interdisciplinary work presented in this paper provides a mathematical structure in the context of evolutionary game theory \cite{frey2010evolutionary, banerjee2015dynamics} and nonlinear control \cite{strogatz2014nonlinear} for rigorous analysis of the social interactions \cite{vega2015biofilms, martin2016laboratory} in both pathogenic and non-pathogenic biofilms. In general, this study yields a framework for analyzing multi-order replicator systems.}\\
\indent$\bullet${ Existing literature primarily addresses the problems on population dynamics with an assumption of $\sum_{i}\tilde{x}_{i}=1$, where $\tilde{x}_{i}$ is the population density of the $i^{th}$ competing strain \cite{hauert2006evolutionary, hauert2008ecological}. However, \textit{non-competing strains} (as an example ``\textit{inactive alleles}'' - metabolic cost for cooperation, production-reception of quorum molecules is zero \cite{czaran2009microbial}) those who do not actively take part in public goods game, can be present in the community \cite{hauert2002replicator}, hence, in this situation $\sum_{i}\tilde{x}_{i}<1$. The current work addresses the existence of non-participating strains in ecological public goods games while analyzing the evolution of the population. Incorporation of non-comp-\\eting strains in the present analysis resembles the actual microbial population in a better way compares to the system with only cooperative and defectives strains \cite{czaran2009microbial}. Moreover, the presence of non-participating strains in the colony of cooperators and defectors gives a scope for rigorous analysis of a second-order replicator system in support of understanding the dynamics of individual competing subpopulations separately for different micro-environmental conditions. Whereas, it could be possible that the first-order replicator system for $\sum_{i=1}^{2}\tilde{x}_{i}=1$ misses some crucial happenings in the population due to the perfectly converse dynamics between competing strains.}\\
\indent$\bullet${ We investigate the stability of replicator equation in general and also with an effect of the spatial pattern. The first analysis presents the global stability of the subpopulation of pathogenic cooperators; hence, it gives a general idea for the adversity in the eradication of pathogenicity. Secondly, analysis with the degree of assortment \cite{van2014density} presents a conditional dominance of cooperators while \textit{non-competing strains} exist in the colony, and this system supports bistability and coexistence \cite{hauert2008ecological}.}\\
\indent$\bullet${ Social interactions in biofilms, between homotypic and heterotypic strains, is one of the leading causes of the failure of contemporary anti-biofilm drugs towards the eradication of bacterial infections \cite{leggett2014war, yurtsev2016oscillatory, kim2016rapid}. We claim that the proposed analysis, based on the game-theoretic interactions between competing strains, can contribute in developing a novel antibiotics, which concerns the eradication of pathogenicity with the regulation of cost of cooperation and degree of assortment - two primary factors involve in the proposed EGT model.}
\section{Model Description}
\label{proposed_model}
A subpopulation of cells (cooperators) in biofilms often secretes costly public goods, whereas non-cooperative strains (defectors) use the produced goods without contributing to the microbial consortium \cite{hibbing2010bacterial}. Here, $x$ and $y$ are considered as the fraction of cooperators and defectors, respectively, and both $x$ and $y$ are non-negative. For a microbial colony of only with the subpopulation of producers and non-producers, $x+y=1$  \cite{wakano2009spatial}. If total population size is $N$, the number of cooperators and defectors are $xN$ and $yN$, respectively. However, in this paper, we consider the colony (with constant population size) comprises not only cooperators and defectors but also has some other strains \cite{hauert2002replicator, czaran2009microbial}, i.e. $x+y<1$. The assumptions we have taken here are that the remaining strains (other than producers and non-producers) in the colony do not contain any public good producers by any means, and the average fitness of the population is primarily due to the fitness of cooperators and defectors as the other strains do not take part (i.e. inactive) in the public good competition \cite{hauert2002replicator, czaran2009microbial}. Therefore, here we consider, non-competing strains do not produce the costly public goods by themselves as well as do not receive the benefit from the public goods produced by the cooperators. In this paper, the stability analysis of the proposed replicator system is discussed in twofold: first, without consideration of assortment level in biofilms as discussed in the following subsection and secondly, Section \ref{analysis_with_segre} presents the analysis with the inclusion of degree of assortment.
\subsection{Population dynamics in biofilms: A replicator system}
\label{pop_dyn}
In this subsection, the steady states of a microbial colony for different micro-environmental conditions are investigated without consideration of spatial pattern. These equilibrium points are calculated from the EGT-based replicator dynamics \cite{frey2010evolutionary, madeo2015game} of cooperators and defectors. The replicator equation \cite{frey2010evolutionary} for $x$ and $y$ with $x+y\neq 1$ can be written as
\begin{align}
\begin{split}
\dot{x}= & x(f_{C}-\bar{f})=x[(1-x)f_{C}-yf_{D}]\\
\dot{y}= & y(f_{D}-\bar{f})=y[(1-y)f_{D}-xf_{C}],
\end{split}
\label{eqn: repli_x_y}
\end{align}
where $f_{C}$ and $f_{D}$ are the average fitness of cooperators and defectors, respectively. $\bar{f}=xf_{C}+yf_{D}$ is the average fitness of the microbial population in ensemble. The standard replicator equation in (\ref{eqn: repli_x_y}) manifests the fitness-dependent growth of a subpopulation. A subpopulation using strategies with a fitness greater than the average fitness ensures increased growth rate while a subpopulation following strategies with lesser fitness than average diminishes in number \cite{frey2010evolutionary}. In the proposed stability analysis, we use this replicator system as a framework because the fitness-dependent growth of a subpopulation also supports the bacterial proliferation in public goods competition in microbial biofilms \cite{frey2010evolutionary}.\\  
\indent The fitness (i.e. proliferation rate) of a competing strain primarily depends on the amount of resource acquired in competition and the metabolic cost of public goods production. Hence, the fitness functions \cite{hauert2008ecological} in (\ref{eqn: repli_x_y}), $f_{C}$ and $f_{D}$ can be defined as
\begin{align}
\begin{split}
f_{C}= \frac{rc}{N}(N-1)x-c\big(1-\frac{r}{N}\big)\;\text{and}\;f_{D}= \frac{rc}{N}(N-1)x,
\end{split}
\label{eqn: def_f_C_f_D}
\end{align}
where $c$ is the metabolic cost of production, $rc$ is the total contribution by a single cooperator to a common pool with multiplication factor $r>1$ (as suggested for ecological public goods games in \cite{hauert2002replicator, hauert2006evolutionary, hauert2008ecological}), $N$ denotes the total population size, and $N>1$. The expression for resource acquired by a center cooperator is $R_{C}=\frac{(N_{C}-1)rc}{N}$, where $N_C$ is the total number of cooperators in the ensemble. The fraction of producers around the center cooperator is $x=\frac{N_{C}-1}{N-1}$, which gives $N_{C}=xN-x+1$. Substituting $N_{C}$ in the expression of $R_{C}$ gives $\frac{rc}{N}(N-1)x$, which is the amount of resource acquired by a center producer from neighboring $(N_{C}-1)$ cooperators. The net metabolic cost of production towards the cooperator is $\sigma= c-\frac{rc}{N}= c(1-\frac{r}{N})$. Therefore, the fitness of a cooperator is the difference between resource acquired and net metabolic cost of cooperation as denoted in the expression of $f_{C}$ in (\ref{eqn: def_f_C_f_D}). The contribution of defectors towards the public goods production is zero, i.e. $\sigma=0$ in the expression of $f_{D}$.\\ 
\indent For $x+y=1$, the solutions of the equation (\ref{eqn: repli_x_y}) are $x=\frac{1}{1+C_{1}^{\prime}e^{\sigma t}}$ and $y=\frac{1}{1+C_{2}^{\prime}e^{-\sigma t}}$, where $C_{1}^{\prime},\;C_{2}^{\prime}$ are the integration constants, and $C_{1}^{\prime},\;C_{2}^{\prime}>0$, therefore, defectors dominate when public goods producers get a lesser return (from its own production) than production cost $(\frac{rc}{N}<c\;\Rightarrow r<N)$. Conversely, pathogenic (to the hosts) cooperators take over the population whenever return exceeds the cost $(r>N)$ \cite{wakano2009spatial}. Hence, for a system with only cooperators and defectors, there is two stable states $(0,1)$ and $(1,0)$ for $r<N$ and $r>N$, respectively. In a control point of view, eradication of infectious biofilm in a system with $x+y=1$ \cite{wakano2009spatial} implies the switching of stable equilibrium from $(1,0)$ to $(0,1)$ (domination of public goods non-producers) by any external influence (drug). Therefore, it might be a bit easier to control such a system with only cooperative and defective strains. On the contrary, in the proposed system with non-participating strains, coexisting with public goods producers and non-producers, we determined $(1,0)$ (domination of public goods producers) as only the global stable equilibrium. This global domination of cooperators states a considerable difficulty in the eradication of pathogenic biofilms compares to the system in literature \cite{hauert2008ecological, wakano2009spatial} as the cooperators predominantly influence the chronic bacterial infections towards the hosts.\\
\indent We consider $A=\frac{rc}{N}(N-1)$ and $B=c(1-\frac{r}{N})$. Hence, the corresponding fitness functions in (\ref{eqn: def_f_C_f_D}) can be redefined in terms of $A$ and $B$ as
\begin{align} 
\begin{split}
f_{C}= Ax-B,\;
f_{D}= Ax.
\end{split}
\label{eqn: def_f_C_f_D_redef}
\end{align}
\indent By substituting (\ref{eqn: def_f_C_f_D_redef}) in (\ref{eqn: repli_x_y}), the modified dynamics is defined as
\begin{align}
\begin{split}
\dot{x}= & x[(Ax-B)(1-x)-Axy]\\
\dot{y}= & y[x(A+B-Ax)-Axy].
\end{split}
\label{eqn: repli_x_y_modi}
\end{align}
\indent Fixed points of the replicator system in (\ref{eqn: repli_x_y_modi}) are calculated using $\dot{x}=0,\;\dot{y}=0$, which manifests the equilibrium points as $(x^{*},y^{*})=(0,0),\;(0,1+\frac{B}{A}),\;(1,0),\;(\frac{B}{A},0)$, and a line of fixed points, $x^{*}+y^{*}=1$. The system in (\ref{eqn: repli_x_y_modi}) contains an infinite number of fixed points in the form $(0,y^{*})$, which are unstable due to the absence of cooperators \cite{doebeli2004evolutionary}. However, we analyze the equilibrium points $(0,0)$, $(0,1+\frac{B}{A})$ and $(0,1)$ as these are directly obtained from (\ref{eqn: repli_x_y_modi}). In the process of evaluating the fixed points from (\ref{eqn: repli_x_y_modi}), a system of equations can be written as 
\begin{align}
\begin{split}
(Ax-B)(1-x)-Axy= & 0\\
x(A+B-Ax)-Axy= & 0.
\end{split}
\label{eqn: fix_pt_imp_rsut}
\end{align} 
\indent By solving (\ref{eqn: fix_pt_imp_rsut}), we can write $B=0$, which implies $r=N$ as $c>0$. Moreover, $B=0$ in (\ref{eqn: fix_pt_imp_rsut}) results $x+y=1$ as $A>0$ and $x\neq 0$ for the system of equations in (\ref{eqn: fix_pt_imp_rsut}). Hence, the microbial ensemble comprises cooperators and defectors only (no non-participating strains) for $r=N$. In other words, the solution of (\ref{eqn: fix_pt_imp_rsut}), i.e. $x+y=1$ is a line of fixed points, which is a special case of (\ref{eqn: repli_x_y_modi}) for $r=N$.
\subsection{Investigation of the stability of equilibrium points using the method of linearization}
\label{sta_inves}
The stability of previously calculated fixed points is investigated for the proposed dynamics as defined in (\ref{eqn: repli_x_y_modi}). The replicator dynamics in (\ref{eqn: repli_x_y_modi}) can be redefined as
\begin{align}
\begin{split}
\dot{x}= & f(x,y)=-Ax^{3}+(A+B)x^{2}-Ax^{2}y-Bx\\
\dot{y}= & g(x,y)=-Axy^{2}-Ax^{2}y+(A+B)xy.
\end{split}
\label{eqn: repli_x_y_fr_stab}
\end{align}
\indent The method of linearization using Jacobian in the neighborhood of equilibrium points is used to investigate the stability of the system with respect to the fixed points. The Jacobian matrix,  $J$ can be expressed from the replicator system in (\ref{eqn: repli_x_y_fr_stab}) as   
\begin{equation}
J= \left[ \begin{array}{cc}
\frac{\partial{f}}{\partial{x}} & \frac{\partial{f}}{\partial{y}} \\
\frac{\partial{g}}{\partial{x}} & \frac{\partial{g}}{\partial{y}}
\end{array} \right],
\label{eqn: jaco}
\end{equation}
where
\begin{align}
\begin{split}
&\frac{\partial{f}}{\partial{x}}=-3Ax^{2}+2(A+B)x-2Axy-B,\;\frac{\partial{f}}{\partial{y}}=-Ax^{2},\\
&\frac{\partial{g}}{\partial{x}}= -Ay^{2}+(A+B)y-2Axy,\;\frac{\partial{g}}{\partial{y}}= -Ax^{2}+(A+B)x-2Axy.
\end{split}
\label{eqn: jaco_val}
\end{align}
\subsubsection{\textbf{Analysis of fixed point} $\mathbf{(x^{*}=\;0,\;y^{*}=\;0)}$}
\label{analysis_0_0}
$(x^{*},y^{*})=\;(0,0)$ signifies the microbial colony solely comprises non-competing strains. The Jacobian matrix, $J$ for (0,0) is calculated using (\ref{eqn: jaco}) and (\ref{eqn: jaco_val}) as  
\begin{equation}
J_{(0,0)}= \left[ \begin{array}{cc}
-B & 0 \\
0 & 0
\end{array} \right].
\label{eqn: jaco_0_0}
\end{equation}
\indent $\tau_{(0,0)}=-B=-c(1-\frac{r}{N})$ and $\Delta_{(0,0)}=0$, where $\tau$ and $\Delta$ denote the trace and determinant of $J$, respectively. For $r<N$, $\tau_{(0,0)}<0$, and for $r>N$, $\tau_{(0,0)}>0$. $\Delta_{(0,0)}=0$ for any relation between $r$ and $N$. ($\tau$, $\Delta$) analysis of the fixed point $(0,0)$ represents the borderline case, hence, the nature of $(0,0)$ cannot be determined directly. Moreover, the eigen values of $J_{(0,0)}$ are $\lambda_{1,2}^{(0,0)}$= ($-B$, $0$), and $y=0$ and $x=0$ are the corresponding eigen vectors. Here, $\lambda_{i}^{(j,k)}$ denotes $i^{\text{th}}$ eigen value of Jacobian, $J_{(j,k)}$ for $(j,k)$ equilibrium point. For $r<N$, $\lambda_{1}^{(0,0)}<0$ and for $r>N$, $\lambda_{1}^{(0,0)}>0$, therefore, the eigen vector $y=0$ converges to $(0,0)$ for $r<N$ and diverges from $(0,0)$ for $r>N$. However, it is difficult to denote the nature (converging or diverging) of eigen vector $x=0$ for $\lambda_{2}^{(0,0)}=0$ from both ($\tau$, $\Delta$) and eigen value analysis of $J_{(0,0)}$. Hence, for further analysis of the nature of eigen vector $x=0$, we use \textit{centre manifold theory} \cite{boldin2006introducing, auger2008aggregation}. A parabolic manifold with centre at (0,0) and tangent to $x=0$ can be defined as $x=h(y)=h_{2}y^{2}$. In the formulation of centre manifold, the replicator dynamics in (\ref{eqn: repli_x_y_fr_stab}) is denoted as
\begin{align}
\begin{split}
\dot{x}=A^{\prime}x+f^{\prime}(x,y)\\
\dot{y}=B^{\prime}y+g^{\prime}(x,y).
\end{split}
\label{eqn: dyna_cen_mani}
\end{align}
\indent By comparing (\ref{eqn: repli_x_y_fr_stab}) and (\ref{eqn: dyna_cen_mani}), we can write $A^{\prime}$= $-B$, $f^{\prime}(x,y)$= $-Ax^{3}+(A+B)x^{2}-Ax^{2}y$, $B^{\prime}=0$ and $g^{\prime}(x,y)$= $-Axy^{2}-Ax^{2}y+(A+B)xy$. We use the following expression \cite{strogatz2014nonlinear} to calculate $h_{2}$ in the equation of centre manifold $h(y)=h_{2}y^{2}$.
\begin{equation}
\dot{h}(y)[B^{\prime}y+g^{\prime}(h(y),y)]-A^{\prime}h(y)-f^{\prime}(h(y),y)=0.
\label{eqn: fr_calc_h_2}
\end{equation} 
\indent By substituting the values of $A^{\prime}$, $B^{\prime}$, $f^{\prime}$ and $g^{\prime}$ in (\ref{eqn: fr_calc_h_2}), we can get
\begin{align}
\begin{split}
& 2h_{2}y[-Ah_{2}^{2}y^{5}-Ah_{2}y^{4}+(A+B)h_{2}y^{3}]+Bh_{2}y^{2}+\\
& [Ah_2^{3}y^{6}+Ah_{2}^{2}y^{5}-(A+B)h_{2}^2y^{4}]=0.
\end{split}
\label{eqn: exp_fr_h_2_0_0}
\end{align}
\indent Neglecting the higher order terms (h.o.ts) $(>\mathcal{O}(y^{2}))$ in (\ref{eqn: exp_fr_h_2_0_0}), we get $Bh_{2}y^{2}=0$ $\Rightarrow h_{2}=0$ as $B,y\neq0$ for the centre manifold. $h_{2}=0$ gives the centre manifold expression as $x=h(y)=h_{2}y^{2}=0$. For $x=0$, we get the system dynamics from (\ref{eqn: repli_x_y_fr_stab}) as $\dot{x}=\dot{y}=0$. Hence, centre manifold $x=0$ is an unstable manifold, and corresponding fixed point (0,0) is an unstable node with respect to the centre manifold $x=0$. In place of $h(y)=h_{2}y^{2}$, if we consider the centre manifold as $x=h(y)=y+h_{2}y^{2}$, we can get the expression after neglecting h.o.ts from (\ref{eqn: fr_calc_h_2}) as $B(y+h_{2}y^{2})=0$ $\Rightarrow h_{2}=-\frac{1}{y}$. $h_{2}=-\frac{1}{y}$ gives the centre manifold equation as $x=0$, and corresponding dynamics $\dot{x}=\dot{y}=0$, which is same as previous. For $r<N$, the eigen vectors $y=0$ converges and $x=0$ diverges, hence, (0,0) can be considered as a saddle-node. But $\lambda_{2}^{(0,0)}=0$ signifies (0,0) is not a saddle for $r<N$, can be considered as an unstable node. For $r>N$, both the eigen vectors diverge from (0,0). Hence, (0,0) is an unstable equilibrium point for both $r<N$ and $r>N$.
\subsubsection{\textbf{Analysis of fixed point} $\mathbf{(x^{*}=\;0,\;y^{*}=\;1+\frac{B}{A})}$}
\label{analysis_0_1+B/A}  
The equilibrium point ($0,1+\frac{B}{A}$) signifies the absence of cooperators in the colony, and a fraction of defectors coexist with the non-participating strains. The Jacobian matrix, $J$ at the equilibrium point ($0,1+\frac{B}{A}$) is $J_{(0,1+\frac{B}{A})}= \left[ \begin{array}{cc}
-B & 0 \\
0 & 0
\end{array} \right]$, which is same as $J_{(0,0)}$. $\tau_{(0,1+\frac{B}{A})}$= $-B=-c(1-\frac{r}{N})$ and $\Delta_{(0,1+\frac{B}{A})}= 0$. ($1+\frac{B}{A}$) can be expressed in terms of $r$ and $N$ as $\bigg[1+\frac{N-r}{r(N-1)}\bigg]$, which is greater than one for $r<N$, hence, $(0,1+\frac{B}{A})$ is only a valid equilibrium point for $r>N$ as $0\leq x,y \leq 1$. For $r>N$, $\tau_{(0,1+\frac{B}{A})}>0$ and $\Delta_{(0,1+\frac{B}{A})}= 0$. Eigen values of $J_{(0,1+\frac{B}{A})}$ are $\lambda_{1,2}^{(0,1+\frac{B}{A})}$= ($-B$, $0$). $\lambda_{1}^{(0,1+\frac{B}{A})}>0$ for $r>N$, hence, corresponding eigen vector $y=0$ diverges from $(0,1+\frac{B}{A})$. To determine the nature of eigen vector $x=0$ for $\lambda_{2}^{(0,1+\frac{B}{A})}=0$, we use centre manifold theory. The corresponding equation of parabolic manifold centered at $(0,1+\frac{B}{A})$ and tangent to $x=0$ is $x=h(y)=h_{2}(y-a)^2$, where $a=(1+\frac{B}{A})$. After neglecting h.o.ts, We can get the expression from (\ref{eqn: fr_calc_h_2}) for $h(y)=h_{2}(y-a)^2$ as 
\begin{align}
\begin{split}
& y^{2}[5Ah_{2}^{3}a^{4}-2Ah_{2}^{2}a^{3}+Bh_{2}]+\\
& y[-4Ah_{2}^{3}a^{5}+Ah_{2}^{2}a^{4}+2(A+B)h_{2}^{2}a^{3}-2Bh_{2}a]+\\
& [Ah_{2}^{3}a^{6}-(A+B)h_{2}^{2}a^{4}+Bh_{2}a^{2}]=0.
\end{split}
\label{eqn: exp_fr_h_2_0_1+B/A}
\end{align}
\indent L.H.S of (\ref{eqn: exp_fr_h_2_0_1+B/A}) is equal to zero if $h_{2}=0$. Hence, the expression of centre manifold $x=h(y)=h_{2}(y-a)^2=0$, and for $x=0$, $\dot{x}=\dot{y}=0$, which manifests diverging eigen vector $x=0$ of $\lambda_{2}^{(0,1+\frac{B}{A})}=0$. For $r>N$, eigen vectors ($y=0$ and $x=0$, respectively) corresponding to the eigen values $(-B, 0)$ diverge from the fixed point, therefore, $(0,1+\frac{B}{A})$ is an unstable node.
\subsubsection{\textbf{Analysis of fixed point} $\mathbf{(x^{*}=\;1,\;y^{*}=\;0)}$}
\label{analysis_1_0}
\begin{figure}[here]
\centering
\begin{tabular}{ll}
 \includegraphics[scale=0.29]{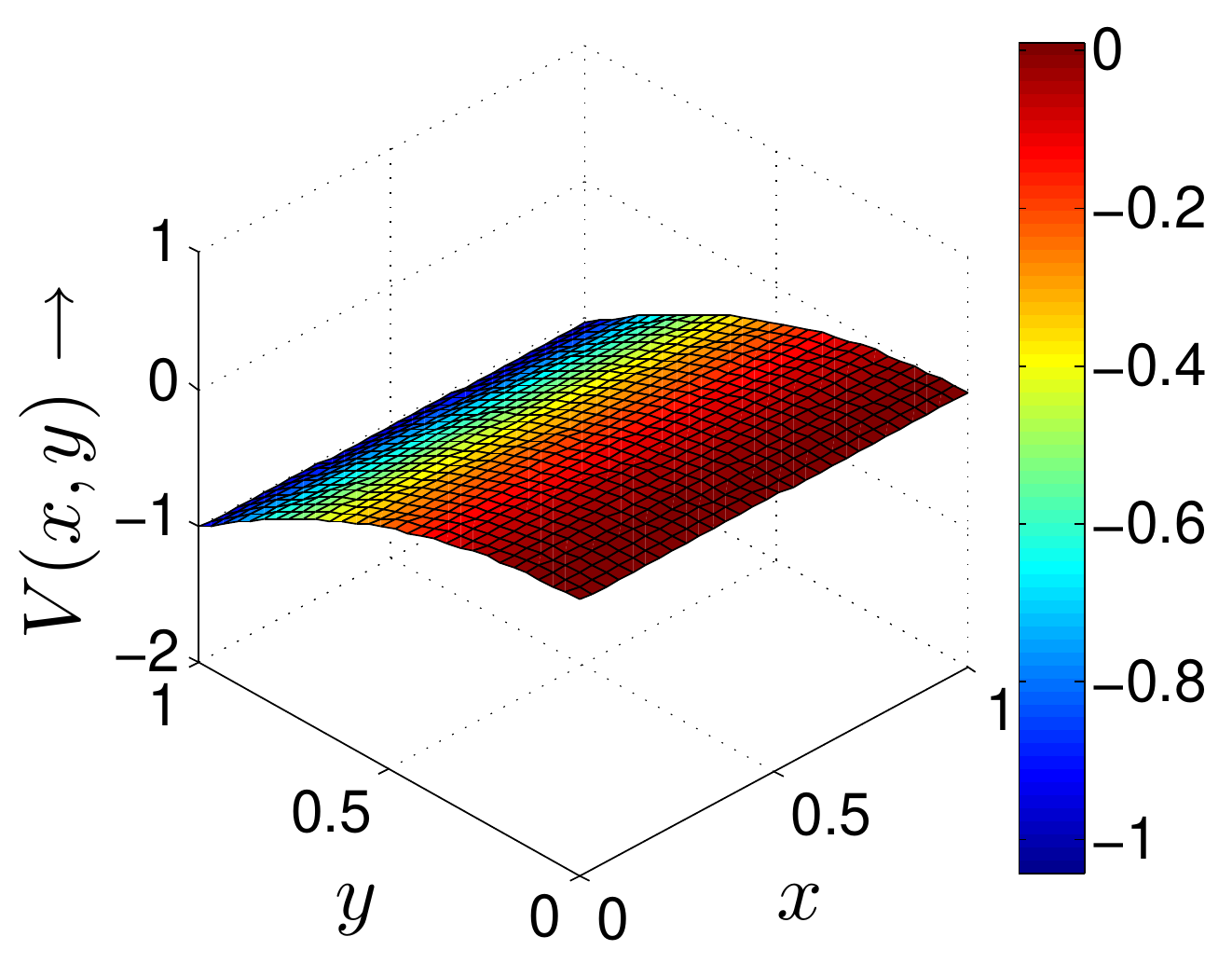} & \includegraphics[scale=0.29]{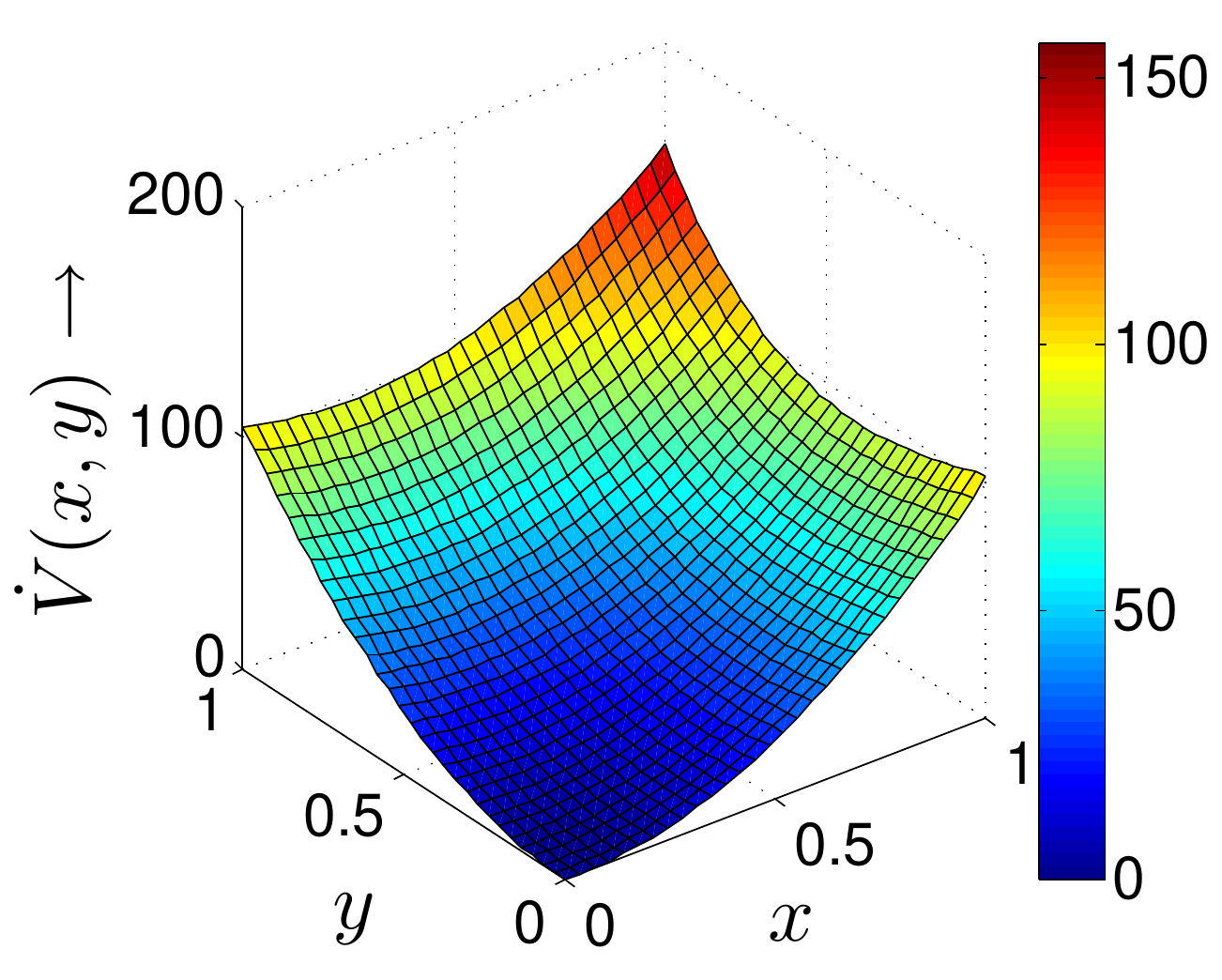}\\
 \;\;\;\;\; \;\;\;\;\;\;\;\;\;\;\;\;\;\;a  &\;\;\;\;\; \;\;\;\;\;\;\;\;\;\;\;\;\;\;b \\
 \includegraphics[scale=0.29]{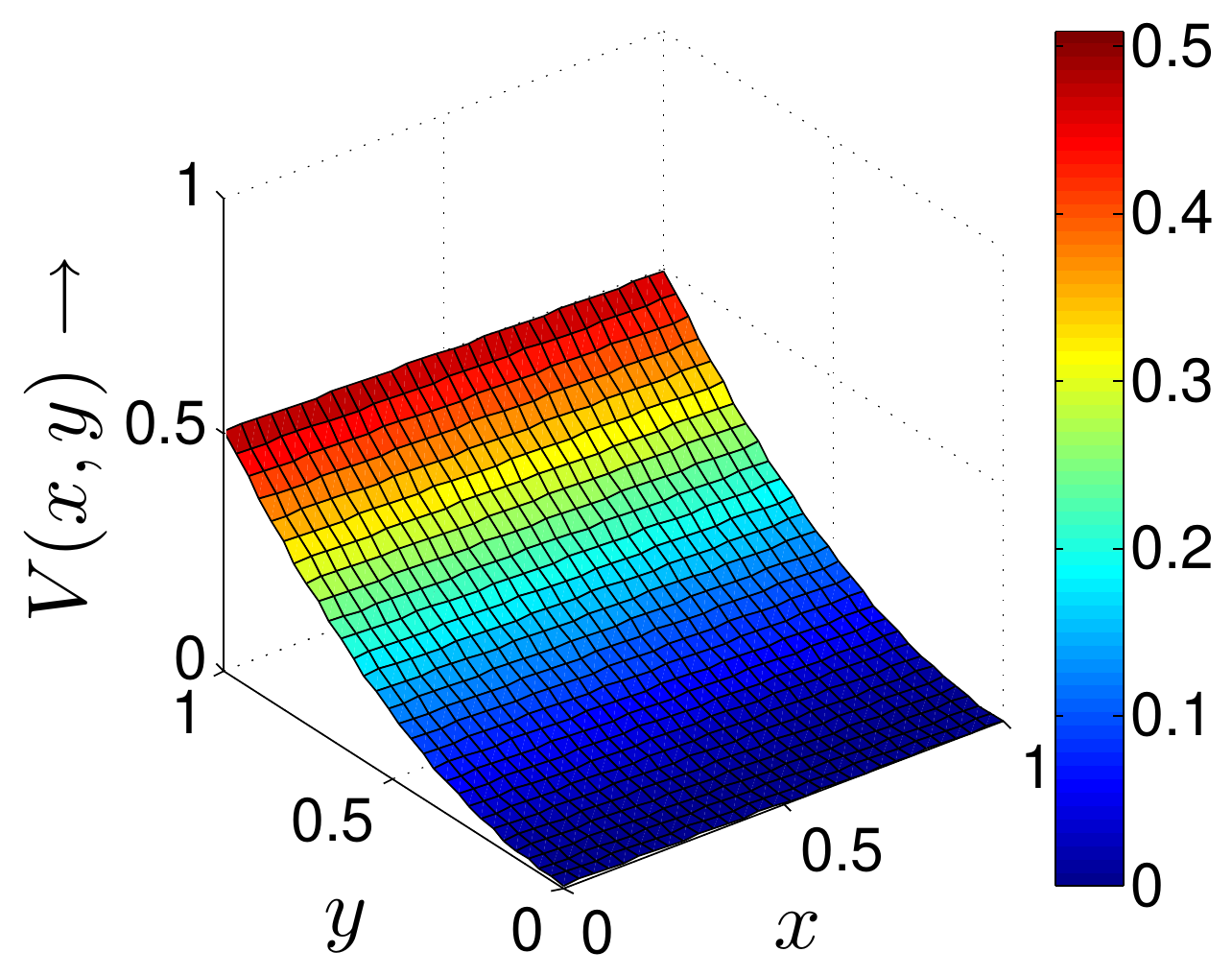}&
\includegraphics[scale=0.29]{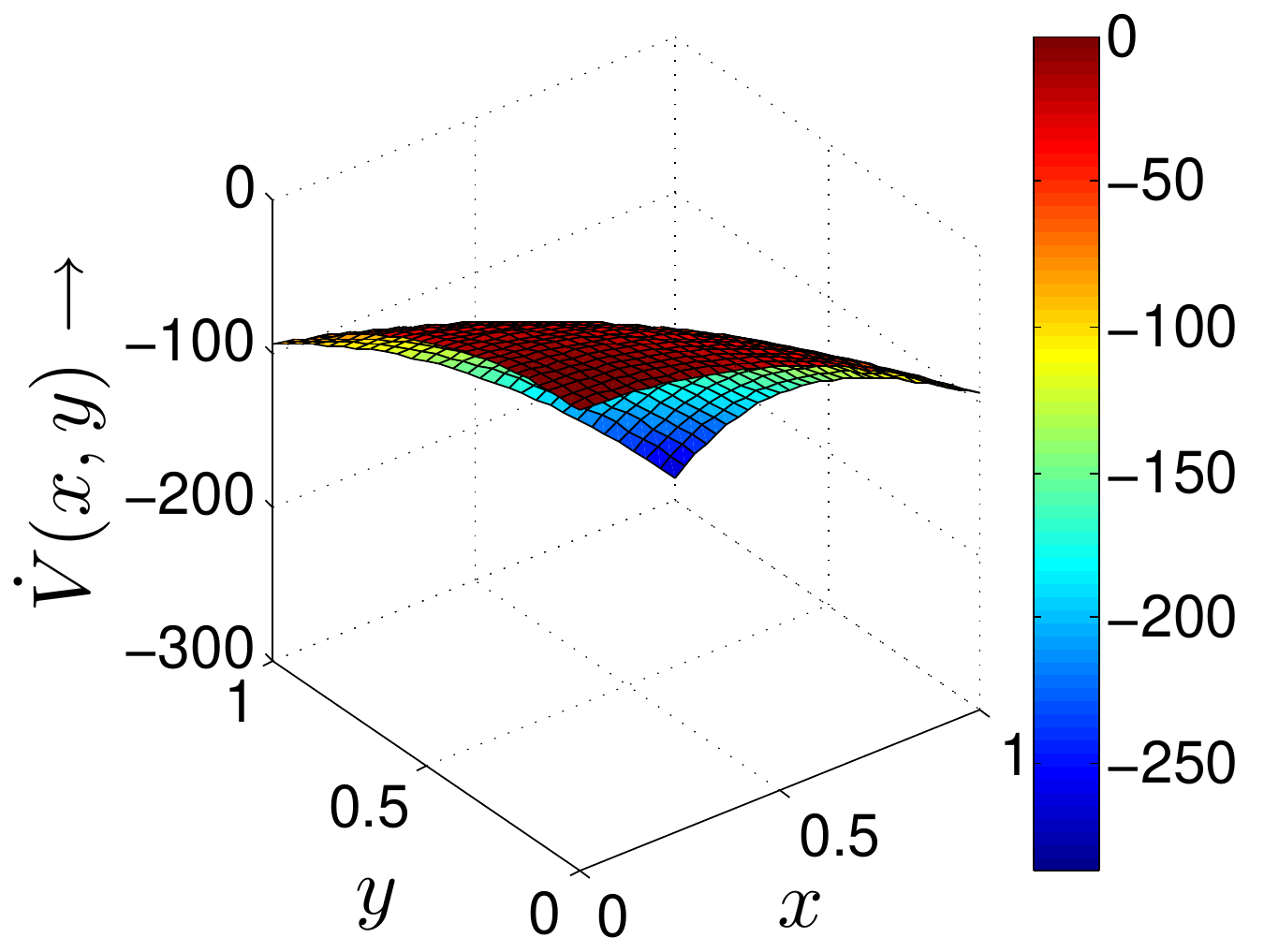} \\
\;\;\;\;\; \;\;\;\;\;\;\;\;\;\;\;\;\;\;c  & \;\;\;\;\; \;\;\;\;\;\;\;\;\;\;\;\;\;\;d 
\end{tabular}
\caption{(Color online) Simulation of the expressions of $V(X)$ and $\dot{V}(X)$ as formulated in (\ref{eqn: V_X}) and (\ref{eqn: V_dot}), respectively for $r<N$ and $r>N$. For $r<N$, (a) $V(X)<0$ and (b) $\dot{V}(X)>0$. For $r>N$, (c) $V(X)>0$ and it is radially unbounded and (d) $\dot{V}(X)<0$}
\label{Fig: r_legt_N_V_V_dot}
\end{figure}

The fixed point $(1,0)$ implies the sole domination of public goods producers in biofilms. The Jacobian matrix, $J$ at the fixed point $(1,0)$ is 
\begin{equation}
J_{(1,0)}= \left[ \begin{array}{cc}
B-A & -A \\
0 & B
\end{array} \right].
\label{eqn: jaco_1_0}
\end{equation}
\indent From (\ref{eqn: jaco_1_0}), we can write $\tau_{(1,0)}=2B-A$ and $\Delta_{(1,0)}=B(B-A)$. $\tau_{(1,0)}=(2B-A)=\frac{c}{N}[N(2-r)-r]<0$ manifests the inequality $\frac{2}{r}<1+\frac{1}{N}$, which is true as $r,N>1$ and $r,N\in \mathbb{Z}$. Hence, $\tau_{(1,0)}<0$ for any relation between $r$ and $N$. $\Delta_{(1,0)}=B(B-A)=\frac{c^{2}}{N}(N-r)(1-r)$, which is $<0$ for $r<N$ and $>0$ for $r>N$. $(\tau_{(1,0)}^{2}-4\Delta_{(1,0)})=A^{2}=[\frac{rc}{N}(N-1)]^{2}$, which is always $>0$. Therefore, from the $(\tau,\Delta)$ analysis, we can ensure that the equilibrium point $(x^{*}=\;1,\;y^{*}=\;0)$ is a saddle point for $r<N$ and a stable node for $r>N$. Moreover, eigen values of $J_{(1,0)}$ are $\lambda_{1,2}^{(1,0)}$= $(B-A,B)$, and corresponding eigen vectors are $y=0$ and $y=-x$. $\lambda_{1}^{(1,0)}= B-A= c(1-r)$, which is $<0$ as $c>0$ and $r>1$. $\lambda_{2}^{(1,0)}= B= c(1-\frac{r}{N})$ $>0$ for $r<N$ and $<0$ for $r>N$. From the eigen value analysis it is clear that the $(1,0)$ is a saddle-node for $r<N$ (as $\lambda_{1}^{(1,0)}<0$, $\lambda_{2}^{(1,0)}>0$), and a stable equilibrium for $r>N$ (as $\lambda_{1}^{(1,0)},\;\lambda_{2}^{(1,0)}<0$). Hence, both $(\tau,\Delta)$ and eigen value analysis of $J_{(1,0)}$ ensure the same nature of $(1,0)$ for $r<N$ and $r>N$. The question remains is that  is $(1,0)$ only stable or asymptotically stable equilibrium point for $r>N$, and if asymptotically stable, then is it locally or globally asymptotically stable? To investigate this question, we analyze the point $(1,0)$ using Lyapunov function candidate \cite{shank2014efficient} as discussed in the following.
\paragraph{\textbf{Global asymptotic stability analysis of} $\mathbf{(x^{*}=\;1, y^{*}=\;0)}$}
\label{asymp_stab_1_0}
The Lyapunov function candidate, $V(X)$ can be defined as 
\begin{equation}
V(X)=X^{T}PX,
\label{eqn: lyapunov_func}
\end{equation}
where the state variable $X$= $[x\;\;\;y]^{T}$, $P$ is a symmetric matrix with principal minors greater than zero, i.e., $P= \left[ \begin{array}{cc}
k_{1} & k_{2} \\
k_{2} & k_{3}
\end{array} \right]$ and $k_{1}, \;(k_{1}k_{3}-k_{2}^{2})>0$. The elements of the matrix $P$ can be determined from the following Lyapunov equation \cite{shank2014efficient}.
\begin{equation}
PJ_{(1,0)}+J_{(1,0)}^{T}P=-Q,
\label{eqn: calc_P}
\end{equation}
where $J_{(1,0)}$ is the Jacobian matrix at the equilibrium point $(1,0)$ as defined in (\ref{eqn: jaco_1_0}), $Q$ is a positive definite matrix, considered here as an identity matrix, hence $Q= \left[ \begin{array}{cc}
1 & 0 \\
0 & 1
\end{array} \right]$.
\indent By solving the Lyapunov equation in (\ref{eqn: calc_P}), we can get the values of $k_{1}$, $k_{2}$ and $k_{3}$ as
\begin{align}
\begin{split}
k_{1}= -\frac{1}{2(B-A)},\;
k_{2}= \frac{k_{1}A}{2B-A},\;
k_{3}= \frac{k_{1}A^{2}}{B(2B-A)}-\frac{1}{2B}.
\end{split}
\label{eqn: k_1_2_3}
\end{align}
\indent Now, the eigen values of $P$ are defined as
\begin{equation}
\lambda_{1,2}^{P}=\frac{\tau_{P}\pm \sqrt{\tau_{P}^{2}-4\Delta_{P}}}{2},
\label{eqn: eigen_val_P}
\end{equation}
where $\tau_{P}=k_{1}+k_{3}=-\frac{1}{2(B-A)}-\frac{A^{2}}{2B(B-A)(2B-A)}-\frac{1}{2B}$, $\Delta_{P}=k_{1}k_{3}-k_{2}^{2}=\Big[\Big(\frac{k_{1}A}{2B-A}\Big)^{2}\Big(\frac{B-A}{B}\Big)\Big]+\frac{1}{4B(B-A)}$.\\ 
\indent $\tau_{P}$ and $\Delta_{P}$ are $>0$ for $r>N$ as $B-A=c(1-r)<0$, $(2B-A)<0$ and for $r>N$, $B<0$. Moreover, $\tau_{P}^{2}-4\Delta_{P}=[(k_{1}-k_{3})^2+4k_{2}^{2}]>0$. As $(\tau_{P}^{2}-4\Delta_{P}),\;\Delta_{P}>0$, we can write $\sqrt{\tau_{P}^{2}-4\Delta_{P}} \in (0,\tau_{P})$, hence, $\lambda_{1,2}^{P}>0$, which implies $P$ is a positive definite matrix. Lyapunov stability theorem suggests that the system is asymptotically stable ($\dot{V}(X)<0$) with respect to the equilibrium point for which the matrix $P$ is positive definite \cite{shank2014efficient}. In the proposed system, $P_{(1,0)}>0$ for $r>N$, hence, at this juncture we can say that the system is locally asymptotically stable with respect to the fixed point $(1,0)$ for $r>N$. By substituting (\ref{eqn: k_1_2_3}) in (\ref{eqn: lyapunov_func}), $V(x,y)$ can be defined (after coordinate transformation) as
\begin{align}
\begin{split}
V(x,y)=  & k_{1}(x-1)^2+2k_{2}(x-1)y+k_{3}y^{2}\\
=  & k_{1}\Big[(x-1)^{2}+\frac{2A}{2B-A}(x-1)y+\frac{A^{2}y^{2}}{B(2B-A)}\Big]-\frac{y^{2}}{2B}.
\end{split}
\label{eqn: V_X}
\end{align}
\indent From (\ref{eqn: V_X}) it is clear that $V(1,0)=0$ and $V(\setminus X=(1,0))>0$ as $k_{1},\;k_{2}(x-1)>0$, and for $r>N$, $\;k_{3}>0$. Hence, the necessary conditions for Lyapunov candidate function are satisfied. $\dot{V}(X)$ can be expressed as
\begin{align}
\begin{split}
\dot{V}(X)= & \Big[\frac{\partial{V}}{\partial{X}}\Big]^{T}\dot{X}\\
= & -2k_{1}Ax^{4}+2k_{1}(2A+B)x^{3}-2k_{1}(A+2B)x^{2}+2k_{1}Bx-\\
&(2k_{1}A+4k_{2}A)x^{3}y+[2k_{1}A+2k_{2}(3A+2B)]x^{2}y-\\
&(4k_{2}A+2k_{3}A)x^{2}y^{2}-(A+2B)xy+\\
&[2k_{2}A+2k_{3}(A+B)]xy^{2}-2k_{3}Axy^{3}.
\end{split}
\label{eqn: V_dot}
\end{align}
\indent The expressions of $V(X)$ and $\dot{V}(X)$ as denoted in (\ref{eqn: V_X}) and (\ref{eqn: V_dot}), respectively are simulated for $r<N$ and $r>N$ in Fig. \ref{Fig: r_legt_N_V_V_dot}. From Fig. \ref{Fig: r_legt_N_V_V_dot} (c) and (d), it is clear that $V(X)$ is radially unbounded \cite{strogatz2014nonlinear} and $\dot{V}(X)<0$ for $r>N$, respectively. Hence, we can conclude that the equilibrium point $(1,0)$ is globally asymptotically stable for $r>N$, whereas $(1,0)$ is not a stable equilibrium for $r<N$ as $V(X)<0$ and $\dot{V}(X)>0$ (Fig. \ref{Fig: r_legt_N_V_V_dot} (a) and (b), respectively). 
\subsubsection{\textbf{Analysis of fixed point} $\mathbf{(x^{*}=\;\frac{B}{A},\;y^{*}=\;0)}$}
\label{analysis_B/A_0}
The equilibrium state $(\frac{B}{A},0)$ represents the absence of defectors in the colony, and cooperators coexist with the non-participating strains. The Jacobian matrix, $J$ at the equilibrium point $(\frac{B}{A},0)$ is
\begin{equation}
J_{(\frac{B}{A},0)}= \left[ \begin{array}{cc}
B(1-\frac{B}{A}) & -\frac{B^{2}}{A} \\
0 & B
\end{array} \right].
\label{eqn: jaco_B/A_0}
\end{equation}
\indent For $r>N$, $\frac{B}{A}=\frac{N-r}{r(N-1)}<0$, hence, $(\frac{B}{A},0)$ is only a valid fixed point for $r<N$. $\tau_{(\frac{B}{A},0)}=2B-\frac{B^{2}}{A}=2c(1-\frac{r}{N})-\frac{c(N-r)^{2}}{rN(N-1)}>0$ implies the inequality $N(2-\frac{1}{r})>1$, which is true as $r,N>1$ and $r\in \mathbb{Z}$, hence, $\tau_{(\frac{B}{A},0)}>0$ for $r<N$. $\Delta_{(\frac{B}{A},0)}=B^{2}(1-\frac{B}{A})=c^{2}(1-\frac{r}{N})^{2}\Big[\frac{N(r-1)}{r(N-1)}\Big]>0$, and $(\tau_{(\frac{B}{A},0)}^{2}-4\Delta_{(\frac{B}{A},0)})=(\frac{B^{2}}{A})^{2}>0$. Therefore, from $(\tau,\Delta)$ analysis, it is clear that the fixed point $(\frac{B}{A},0)$ is an unstable node for $r<N$. Moreover, eigen values of $J_{(\frac{B}{A},0)}$ are $\lambda_{1,2}^{(\frac{B}{A},0)}=(B(1-\frac{B}{A}),\;B)$. For $r<N$, both $\lambda_{1}^{(\frac{B}{A},0)}=B(1-\frac{B}{A})=\frac{c(N-r)(r-1)}{r(N-1)}$ and $\lambda_{2}^{(\frac{B}{A},0)}=B=c(1-\frac{r}{N})$ are $>0$, hence, eigen value analysis also ensured that $(\frac{B}{A},0)$ is an unstable equilibrium for $r<N$. 
\subsubsection{\textbf{Analysis of fixed point} $\mathbf{(x^{*}=\;0,\;y^{*}=\;1)}$}
\label{analysis_0_1} 
\begin{figure}[here]
\centering
\begin{tabular}{ll}
\includegraphics[scale=0.345]{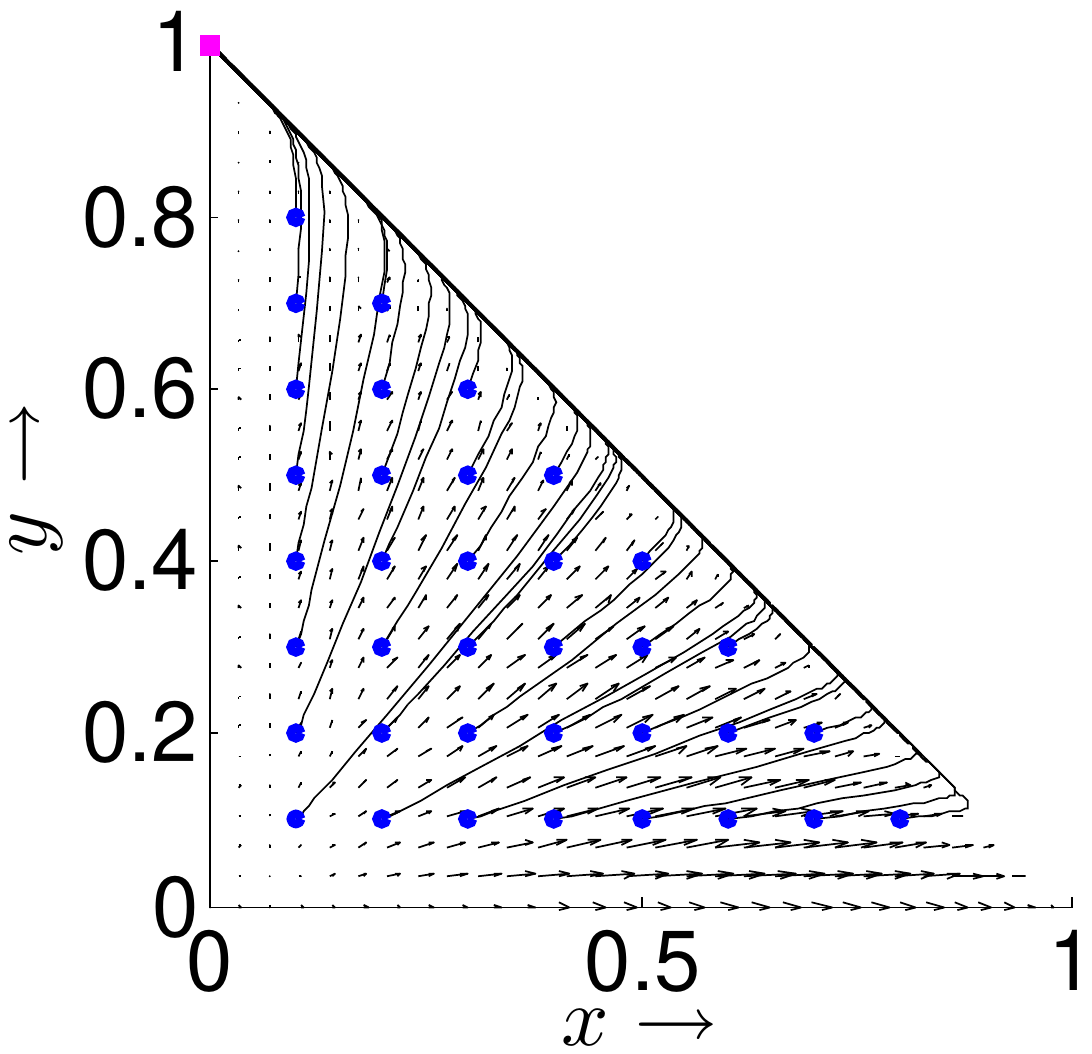} & \includegraphics[scale=0.345]{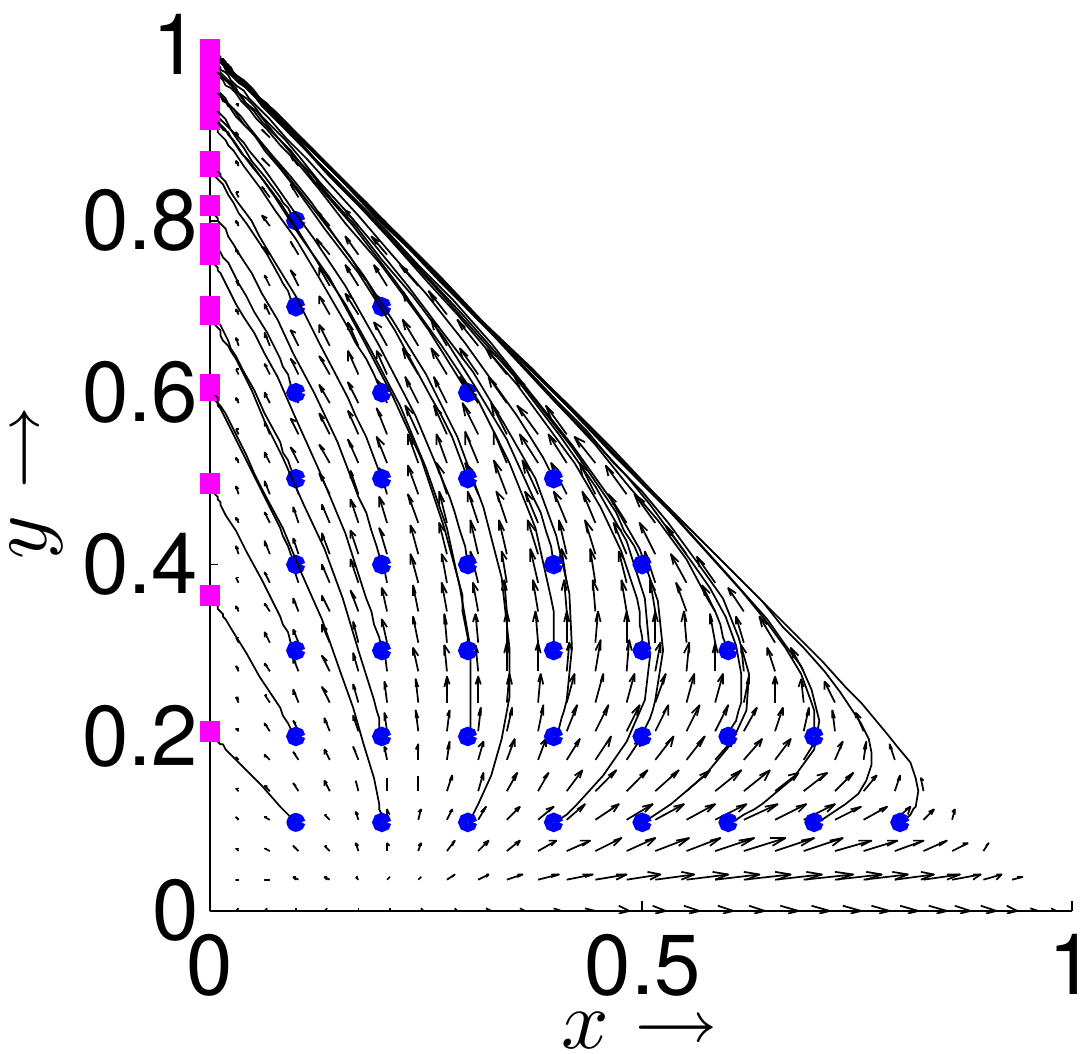}\\
\;\;\;\;\; \;\;\;\;\;\;\;\;\;\;\;\;\;\;\;\;\;\;\;a  &\;\;\;\;\; \;\;\;\;\;\;\;\;\;\;\;\;\;\;\;\;\;\;\;b \\
 \includegraphics[scale=0.345]{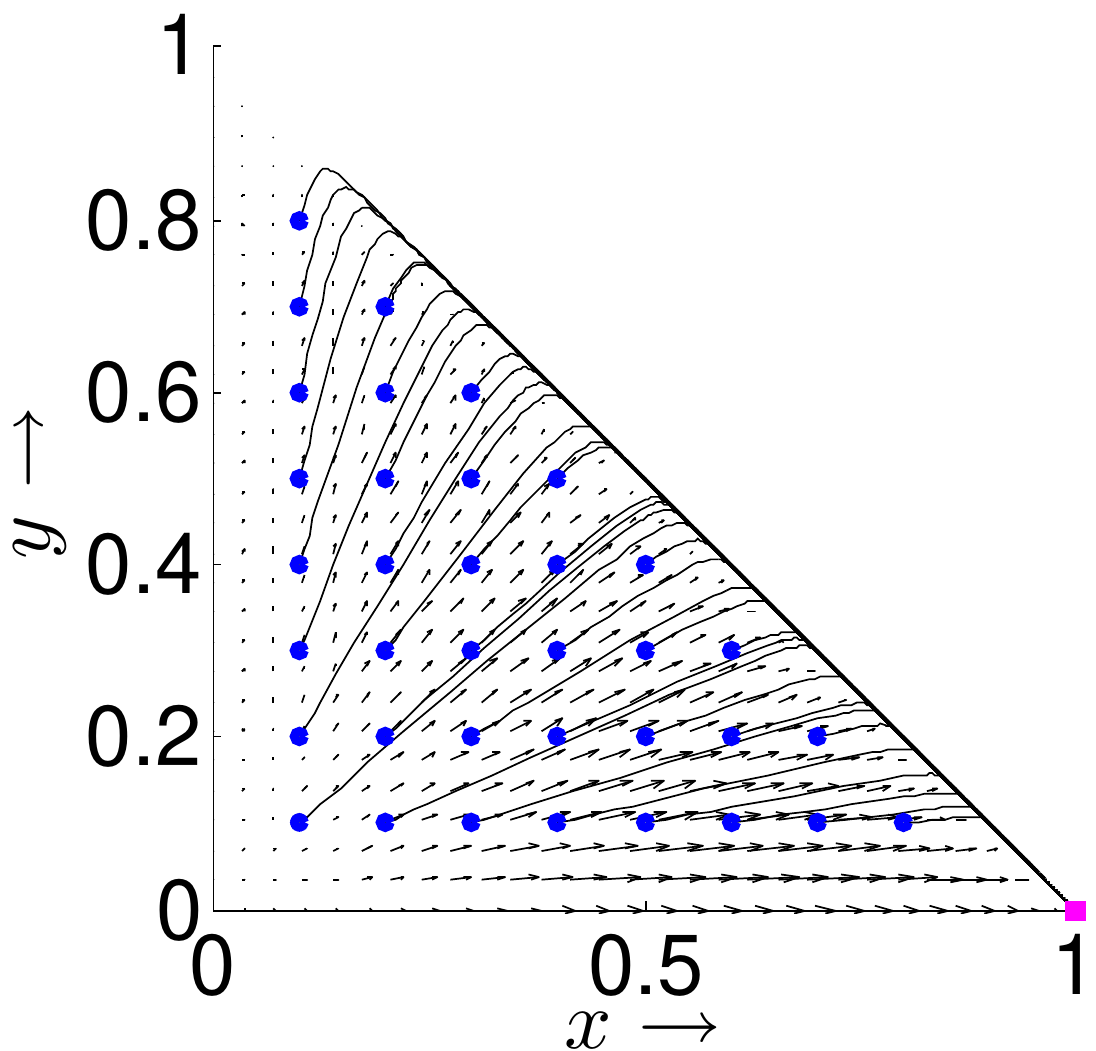}&
\includegraphics[scale=0.345]{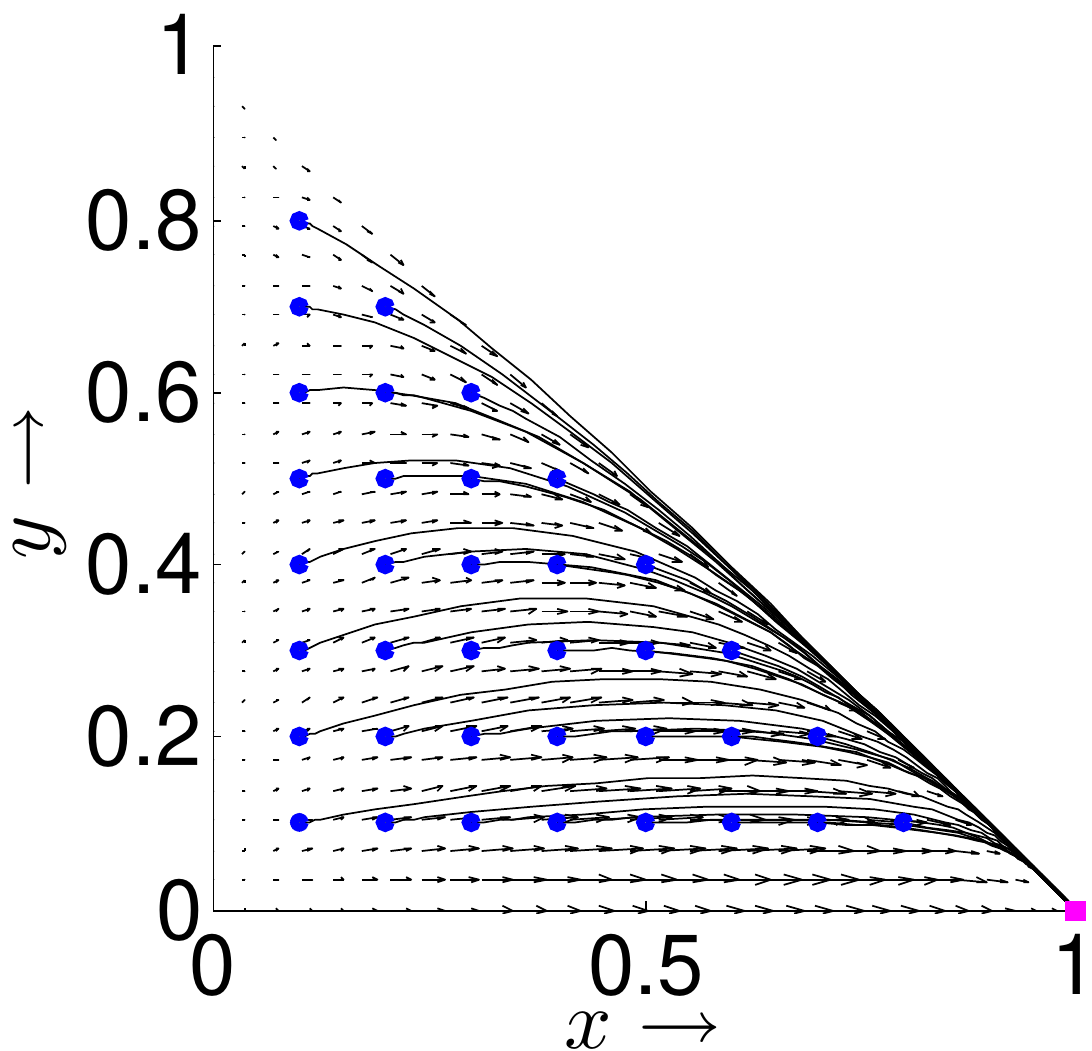} \\
\;\;\;\;\; \;\;\;\;\;\;\;\;\;\;\;\;\;\;\;\;\;\;\;c  & \;\;\;\;\; \;\;\;\;\;\;\;\;\;\;\;\;\;\;\;\;\;\;\;d 
\end{tabular}
\caption{(Color online) Phase portraits for (a) $r<N$, (b) $r<<N$, (c) $r>N$ and (d) $r>>N$. \tikzcircle[fill=blue]{4.3pt} and \crule[magenta]{0.3cm}{0.3cm} denote initial conditions and equilibrium point(s), respectively.}
\label{Fig: phase_potrait_r_legt_N}
\end{figure}

\begin{figure}[here]
\centering
\begin{tabular}{ll}
\includegraphics[scale=0.28]{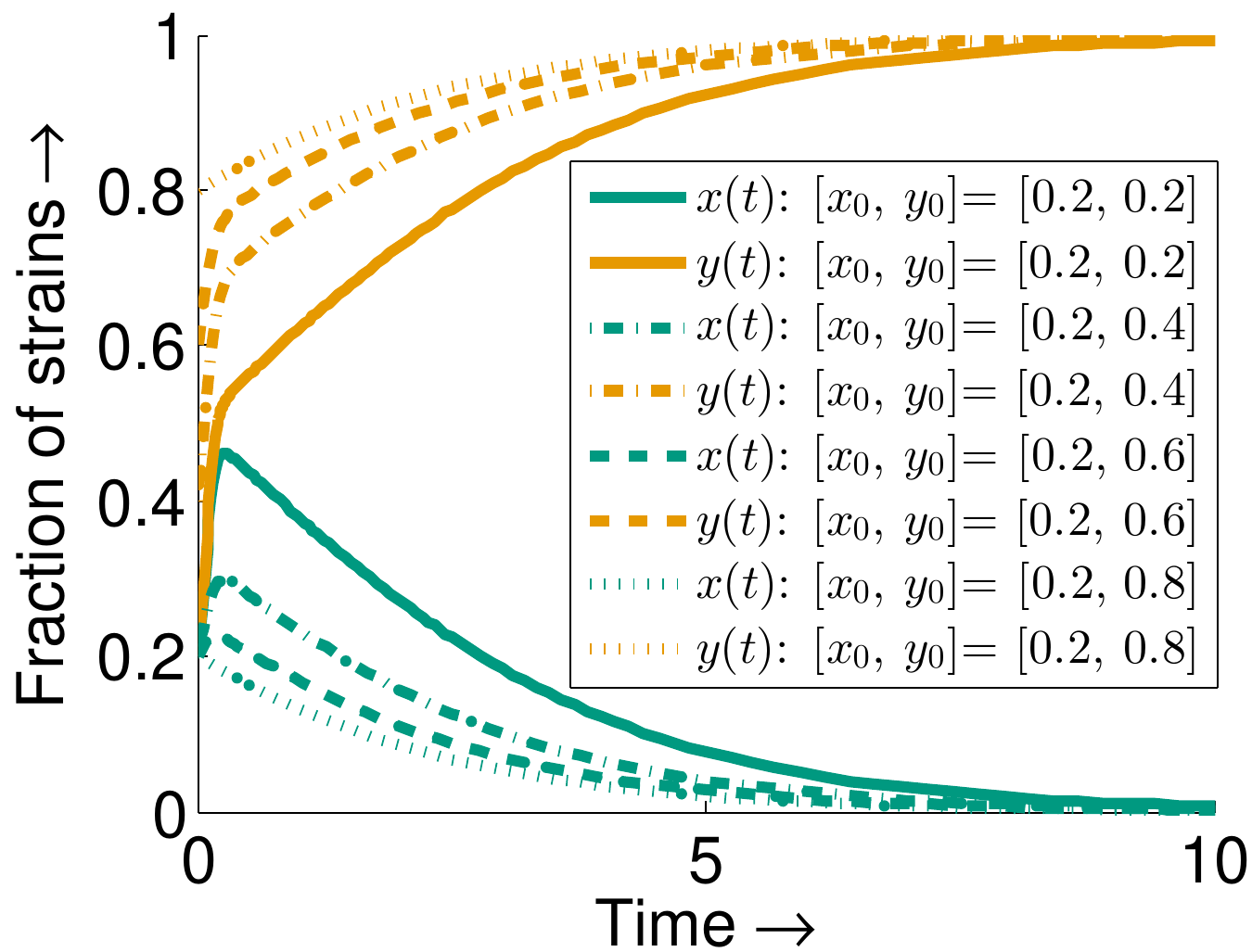} &  \includegraphics[scale=0.28]{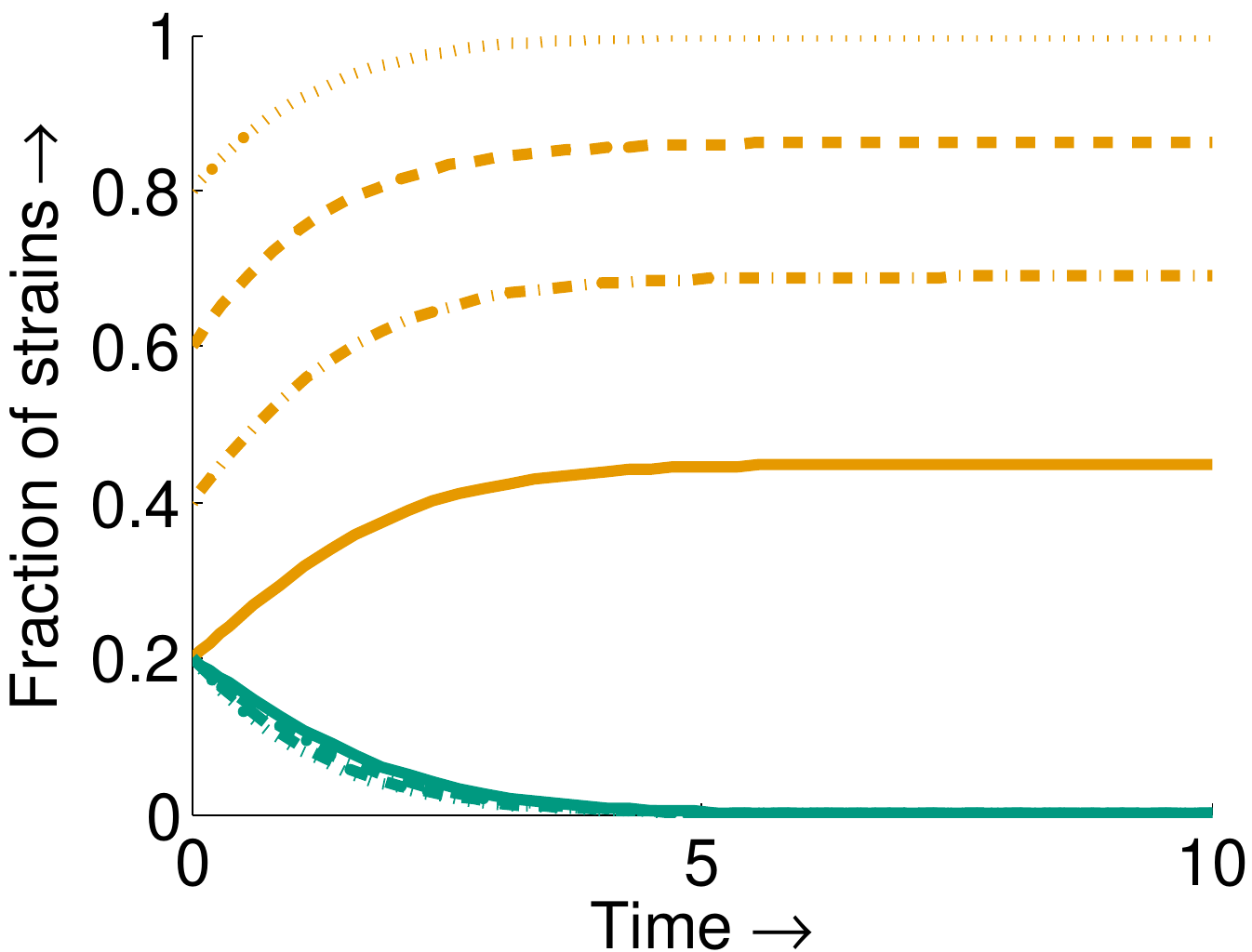}\\
\;\;\;\;\;\;\;\;\;\;\;\;\;\;\;\;\;\;\;\;\;\;\;\;a  & \;\;\;\;\;\;\;\;\;\;\;\;\;\;\;\;\;\;\;\;\;\;\;\;b\\
\includegraphics[scale=0.28]{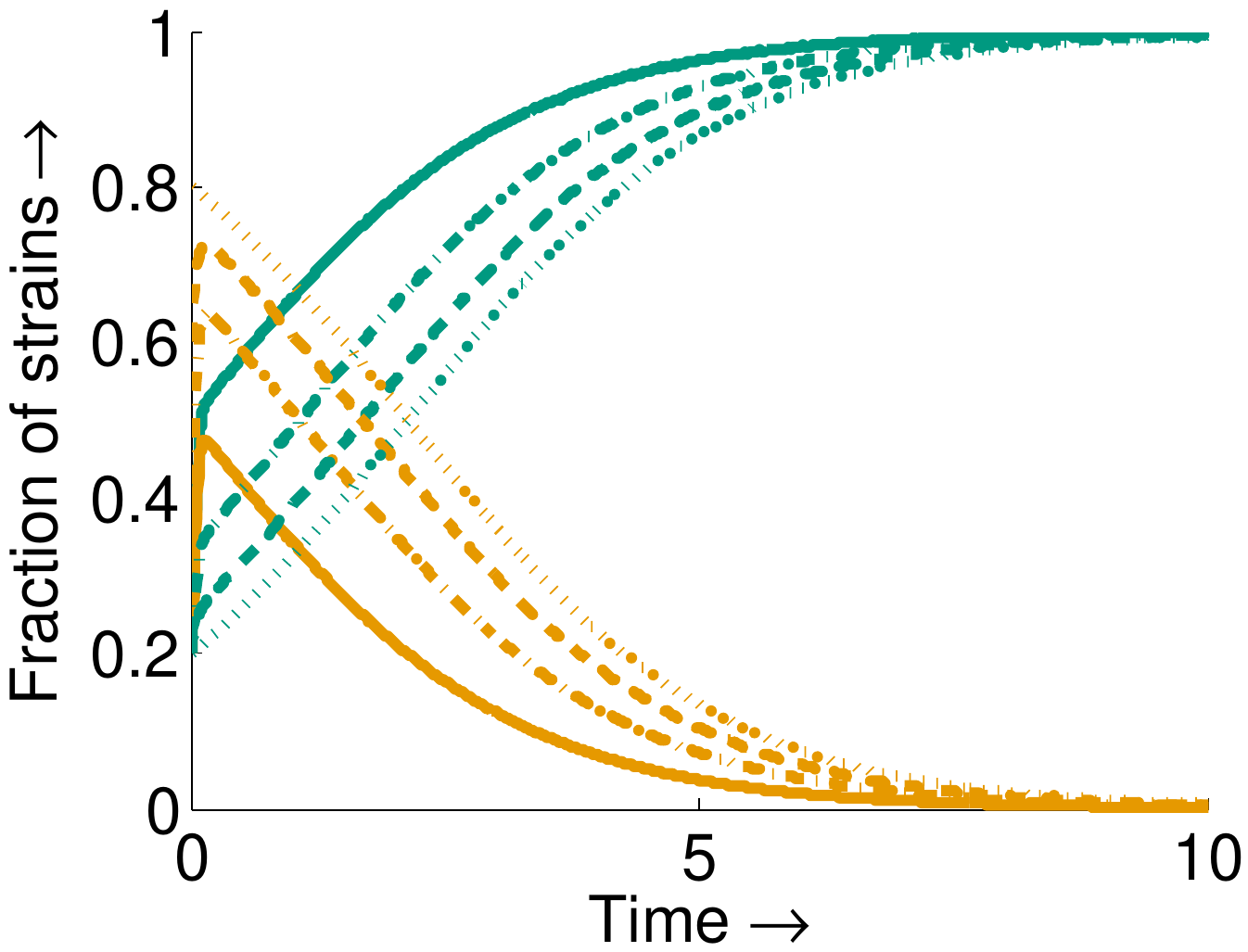} &  \includegraphics[scale=0.28]{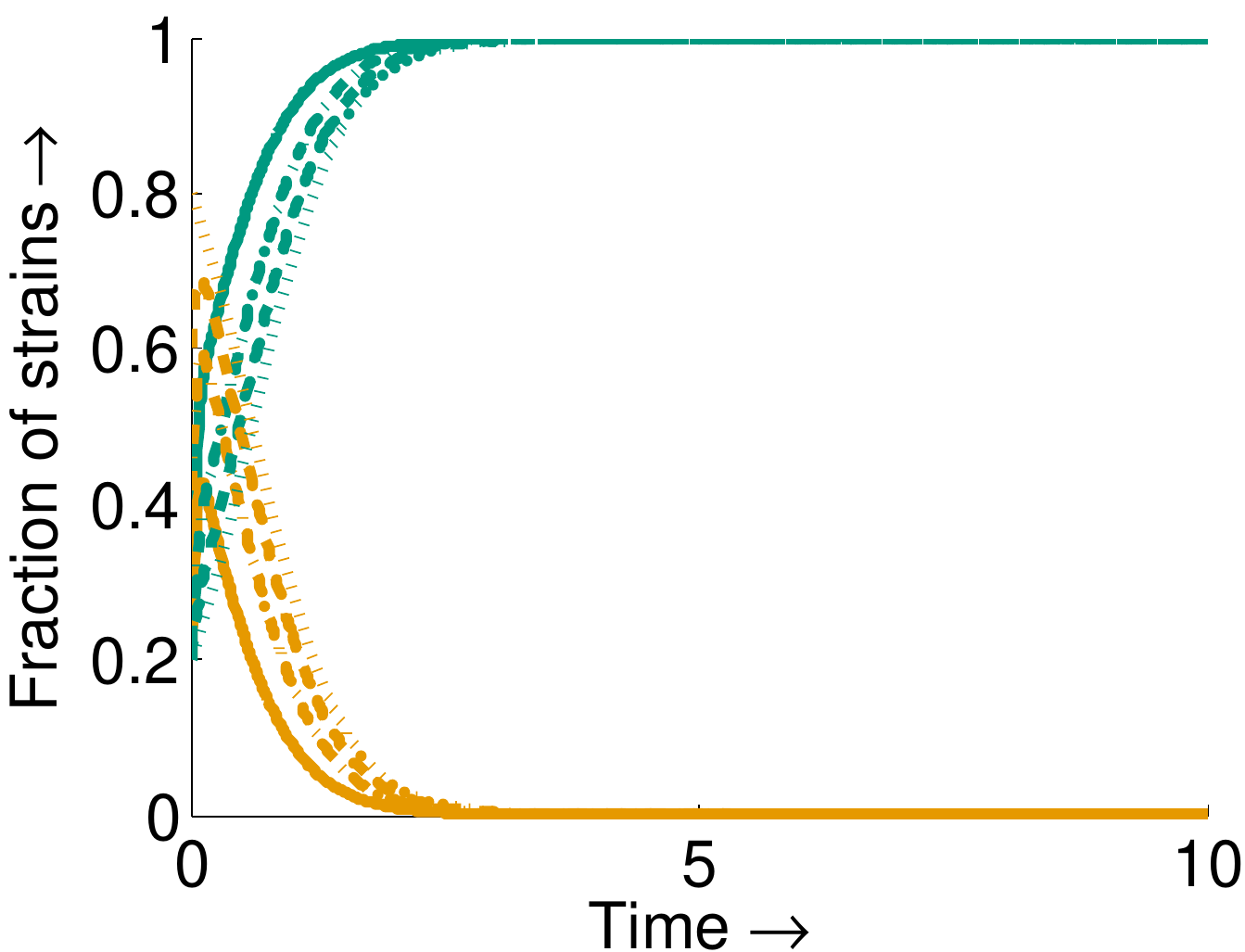}\\
\;\;\;\;\;\;\;\;\;\;\;\;\;\;\;\;\;\;\;\;\;\;\;\;c  & \;\;\;\;\;\;\;\;\;\;\;\;\;\;\;\;\;\;\;\;\;\;\;\;d
\end{tabular}
\caption{(Color online) Simulation of dynamics of $x$ and $y$ from (\ref{eqn: repli_x_y_fr_stab}). (a) ($x^{*}$, $y^{*}$)= (0,1) is the stable equilibrium for $r<N$, (b) (0,1) is an unstable fixed point for $r<<N$ as the fraction of defectors converge to different steady states depending on the initial conditions. ($x^{*}$, $y^{*}$)= (1,0) is the stable fixed point for (c) $r>N$ and (d) $r>>N$. Legend of figure (a) is applicable to the figures (b), (c) and (d).}
\label{Fig: dynamics_r_less_N_r_grt_N}
\end{figure}

\begin{figure}[here]
\centering
\begin{tabular}{ll}
\includegraphics[scale=0.34]{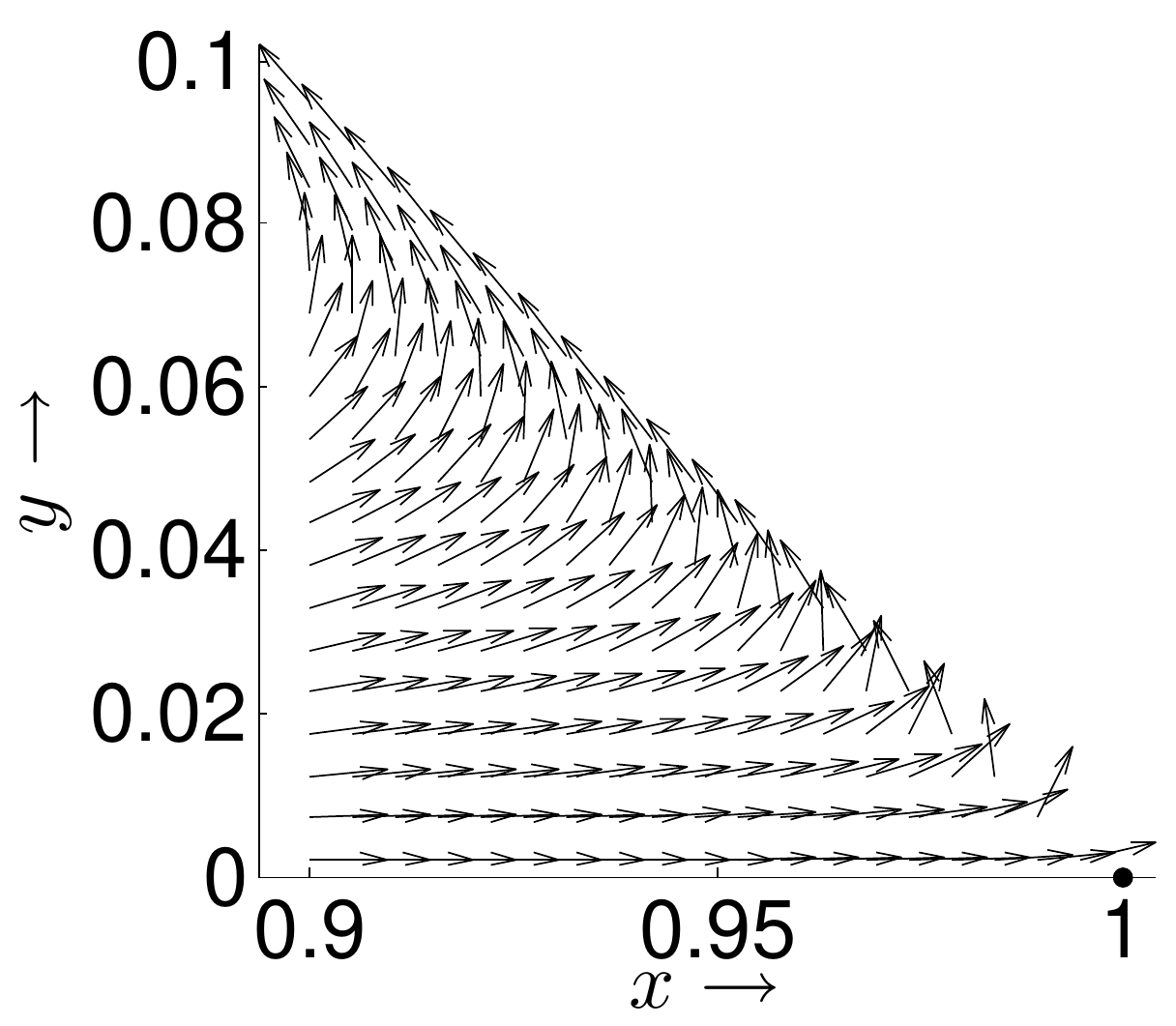} & \includegraphics[scale=0.34]{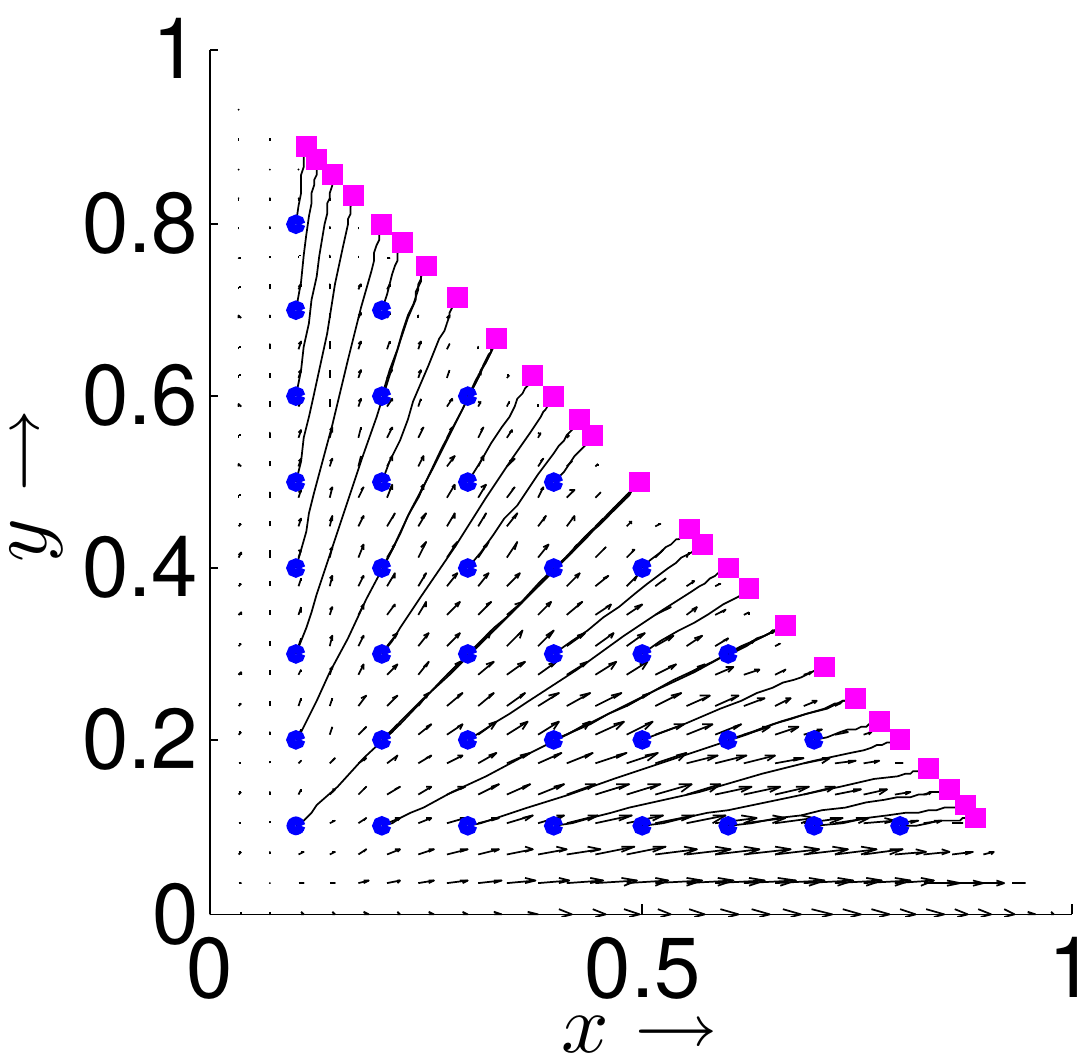}\\
\;\;\;\;\; \;\;\;\;\;\;\;\;\;\;\;\;\;\;\;\;\;\;\;\;\;\;\;a  & \;\;\;\;\;\;\;\;\;\;\;\;\;\;\;\;\;\;\;\;\;\;\;\;\;b 
\end{tabular}
\caption{(Color online) (a) The equilibrium point $(1,0)$ is a saddle for $r<N$, (b) proposed dynamical system in (\ref{eqn: repli_x_y_fr_stab}) converges to the line $x+y=1$ for $r=N$. \tikzcircle[fill=blue]{4.3pt} and \crule[magenta]{0.3cm}{0.3cm} denote initial conditions and equilibrium points, respectively.}
\label{Fig: phase_potrait_r_le_N_saddle_r_eq_N}
\end{figure}

The fixed state $(0,1)$ suggests the `full dominance of defectors' in the colony. We analyze the fixed point $(0,1)$ as it is one of the boundary solutions to $x^{*}+y^{*}=1$. The Jacobian matrix, $J$ at $(0,1)$ is
\begin{equation}
J_{(0,1)}= \left[ \begin{array}{cc}
-B & 0 \\
B & 0
\end{array} \right].
\label{eqn: jaco_0_1}
\end{equation}
\indent Eigen values of $J_{(0,1)}$ are $\lambda_{1,2}^{(0,1)}=(-B,\;0)$, and corresponding eigen vectors are $y=-x$ and $x=0$. $\lambda_{1}^{(0,1)}=-B=-c(1-\frac{r}{N})<0$ for $r<N$ and $>0$ for $r>N$. Hence, the eigen vector $y=-x$ converges to $(0,1)$ for $r<N$, and diverges from the fixed point for $r>N$. Next, we investigate the nature of eigen vector $x=0$ corresponding to eigen value $\lambda_{2}^{(0,1)}=0$ using centre manifold theory as discussed in subsections \ref{analysis_0_0} and \ref{analysis_0_1+B/A}.\\ 
\indent The parabolic manifold centered at $(0,1)$ and tangent to $x=0$ can be considered as $x=h(y)=h_{2}(y-1)^{2}$. To calculate the value of $h_{2}$,   
we use equation (\ref{eqn: fr_calc_h_2}), and corresponding expression after neglecting h.o.ts is
\begin{align}
\begin{split}
& y^{2}[5Ah_{2}^{3}-2Ah_{2}^{2}+Bh_{2}]+y[-4Ah_{2}^{3}+3Ah_{2}^{2}+2Bh_{2}^{2}-2Bh_{2}]+ \\
& [Ah_{2}^{3}-(A+B)h_{2}^{2}+Bh_{2}]=0.
\end{split}
\label{eqn: exp_fr_h_2_0_1}
\end{align}
\indent Equation (\ref{eqn: exp_fr_h_2_0_1}) resembles (\ref{eqn: exp_fr_h_2_0_1+B/A}) with $a=1$. The solution to (\ref{eqn: exp_fr_h_2_0_1}) is $h_{2}=0$, which gives the centre manifold as $x=0$, and corresponding system dynamics is $\dot{x}=\dot{y}=0$. Therefore, $x=0$ is an unstable manifold, and diverges from the equilibrium point $(0,1)$. Hence, for both $r<N$ and $r>N$, $(0,1)$ is an unstable fixed point.\\
\indent The summary of the stability analysis of fixed points (as discussed in subsections \ref{analysis_0_0} to \ref{analysis_0_1}) is tabulated in Table \ref{tab_summary}.
\begin{table}[here]
\begin{center}
\caption{Summary of stability analysis of fixed points}
\begin{tabular}{cccc}
\textbf{Fixed points} & \multicolumn{2}{c}{\textbf{Nature of fixed points}}\\
\hline\noalign{\smallskip}
& $r<N$ & $r>N$\\
\hline\noalign{\smallskip}
(0,0) & Unstable node & Unstable node\\
$(0,1+\frac{B}{A})$ & Not valid & Unstable node\\
\textbf{(1,0)} & \textbf{Saddle-node} & \begin{tabular}{@{}c@{}}\textbf{Globally asymptotically}\\\textbf{stable node}\end{tabular} \\
$(\frac{B}{A},0)$ & Unstable node & Not valid\\
(0,1) & Unstable node & Unstable node \\
\hline\noalign{\smallskip}
\end{tabular}
\label{tab_summary} 
\end{center}
\end{table}
\\\indent Phase portraits for $r<N$ and $r>N$ are simulated in Fig. \ref{Fig: phase_potrait_r_legt_N}. For this simulation, we consider the proposed dynamics as formulated in (\ref{eqn: repli_x_y_fr_stab}). Though for $r<N$, all the initial conditions converge to the equilibrium point $(0,1)$ (figure (a)), for $r<<N$ the stability of $(0,1)$ does not hold true as depicted in figure (b). Hence, for $r<N$, $(0,1)$ is an unstable equilibrium. Whereas figures (c) and (d) confirm $(1,0)$ as globally asymptotically stable equilibrium for both $r>N$ and $r>>N$, hence, $(1,0)$ is only the globally asymptotically stable fixed point for the proposed dynamical system. The dynamics of the fraction of cooperators and defectors are simulated from (\ref{eqn: repli_x_y_fr_stab}), and the results are depicted in Fig. \ref{Fig: dynamics_r_less_N_r_grt_N}. Fig. \ref{Fig: dynamics_r_less_N_r_grt_N} clearly states that $(1,0)$ is globally stable for $r>N$ ((c) and (d)), whereas $(0,1)$ is an unstable node for $r<N$ ((a) and (b)). The simulation results in Fig. \ref{Fig: phase_potrait_r_legt_N} and Fig. \ref{Fig: dynamics_r_less_N_r_grt_N} match with the stability analysis of equilibrium points as discussed in subsections \ref{analysis_0_0} to \ref{analysis_0_1}.\\
\indent Although $(1,0)$ is a globally asymptotically stable equilibrium for $r>N$, for $r<N$, the fixed point $(1,0)$ is a saddle-node as discussed in the subsection \ref{analysis_1_0}, and corresponding phase portrait (zoomed in the neighborhood of $(1,0)$) is depicted in Fig. \ref{Fig: phase_potrait_r_le_N_saddle_r_eq_N} (a). For $r=N$, the proposed dynamics in (\ref{eqn: repli_x_y_fr_stab}) converges to $x+y=1$, i.e. the microbial colony consists of cooperators and defectors only. The simulated phase portrait for $r=N$ is shown in Fig. \ref{Fig: phase_potrait_r_le_N_saddle_r_eq_N} (b) which ensures the convergence of the system towards $x+y=1$.
\subsection{Inference from the above analysis}
\label{infer}
From the previous stability analysis, it is clear that the fixed point $(1,0)$ is only a globally asymptotically stable equilibrium for $r>N$ for the proposed dynamics in (\ref{eqn: repli_x_y_fr_stab}). The equilibrium point $(1,0)$ signifies that the microbial colony is dominated by the public goods producers (so called cooperators), which are primarily responsible for the persistence of microbial infections \cite{bjarnsholt2013role, nadell2015extracellular}. Due to the asymptotically stable nature of the equilibrium point $(1,0)$, the microbial system has an inherent tendency to occupy the whole consortium by the pathogenic cooperators, and this is the reason for the difficulty in the eradication of infectious biofilms. 
\section{Analysis of Replicator Dynamics With Spatial Pattern}
\label{analysis_with_segre}
\begin{figure}[here]
\centering
\begin{tabular}{ll}
\includegraphics[scale=0.28]{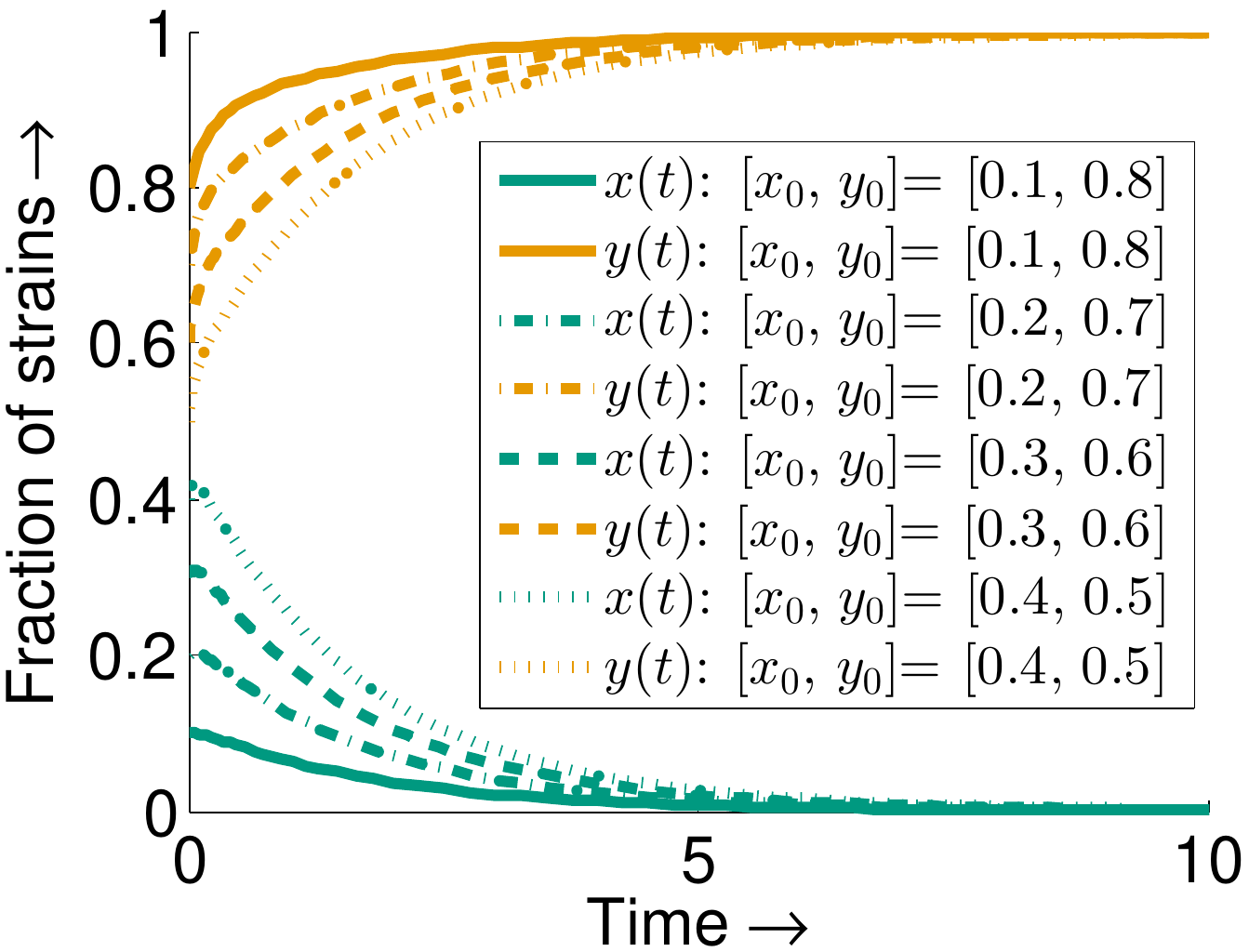} &  \includegraphics[scale=0.28]{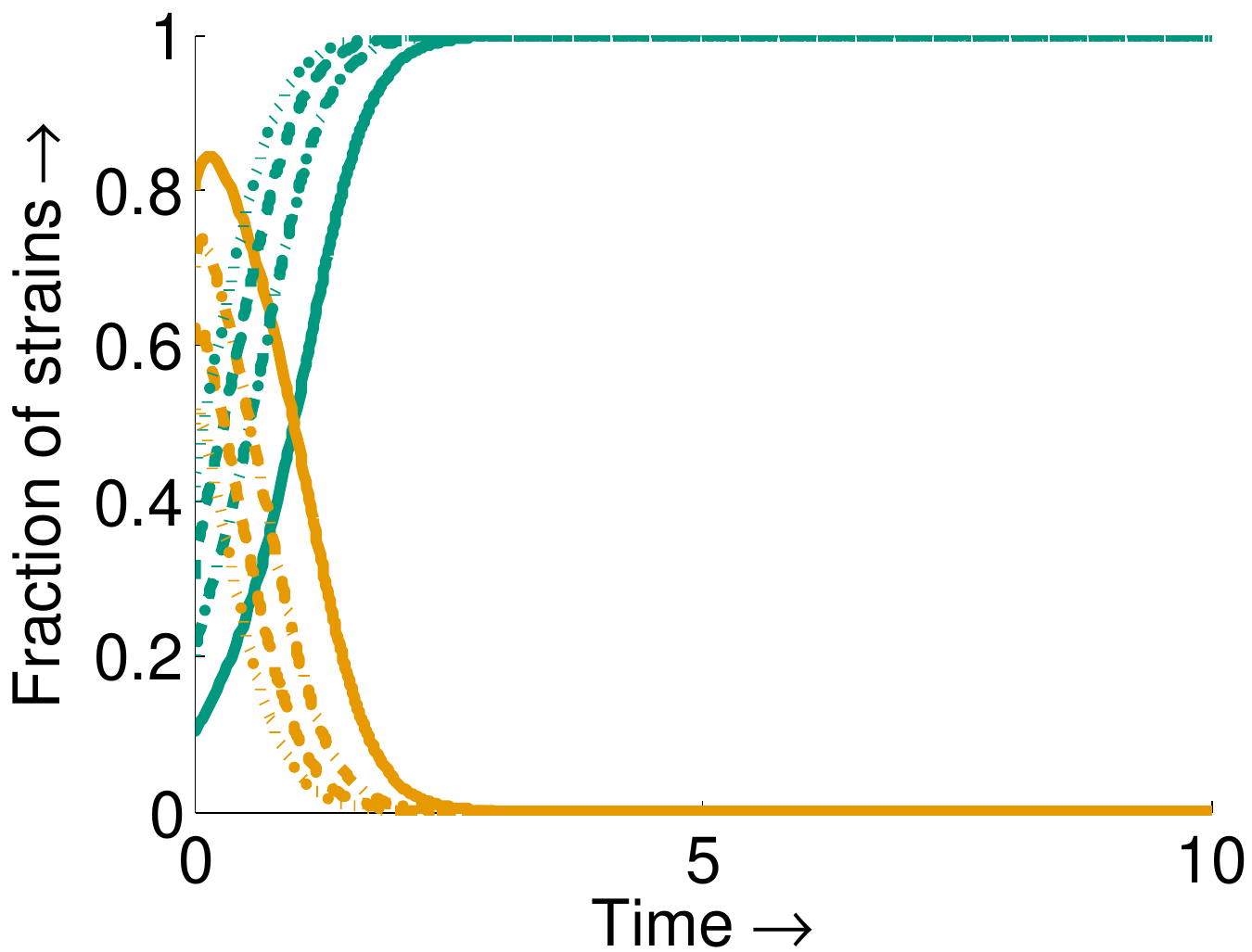}\\
\;\;\;\;\;\;\;\;\;\;\;\;\;\;\;\;\;\;\;\;\;\;\;\;a  & \;\;\;\;\;\;\;\;\;\;\;\;\;\;\;\;\;\;\;\;\;\;\;\;b\\
\end{tabular}
\begin{tabular}{ll}
\includegraphics[scale=0.28]{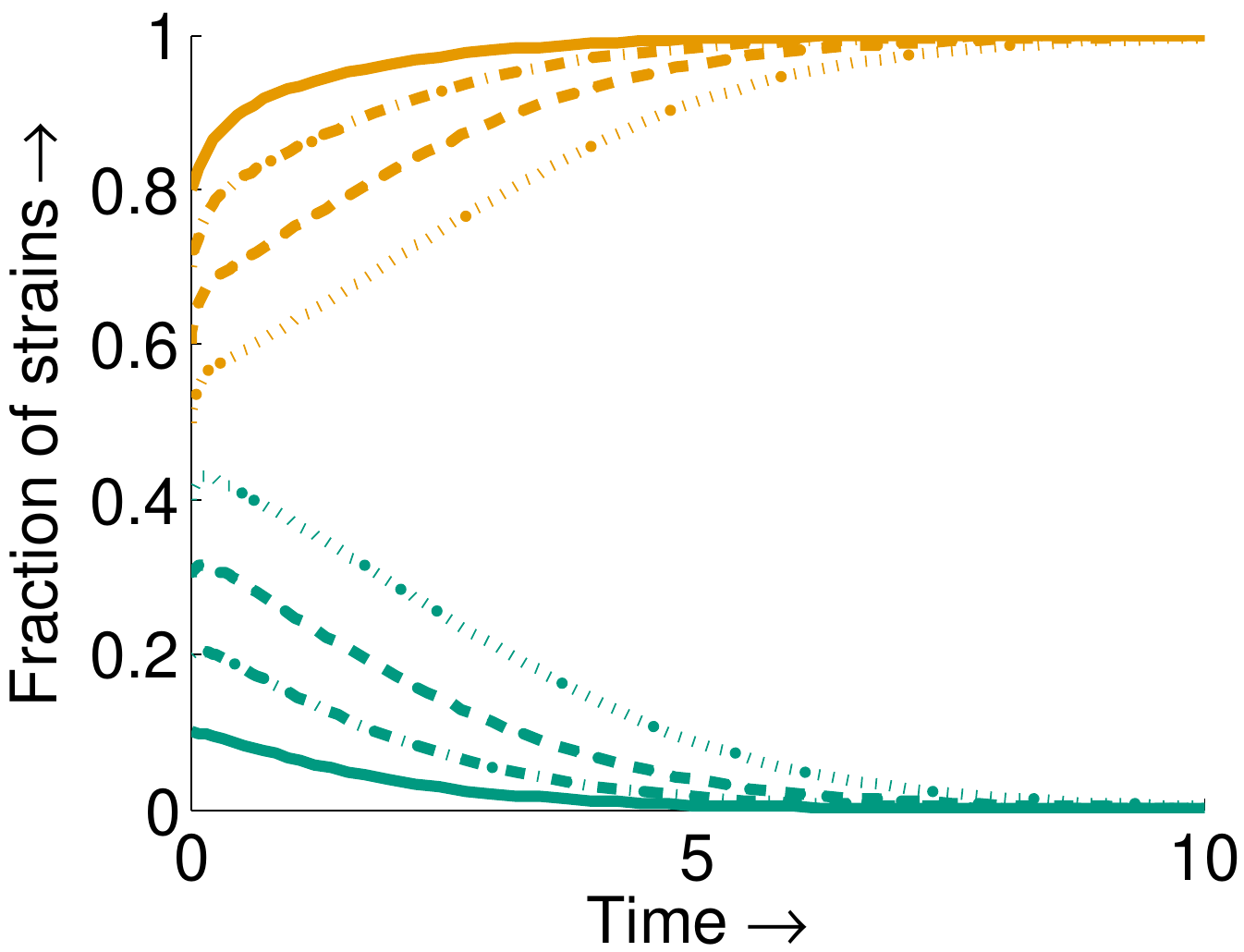} &  \includegraphics[scale=0.28]{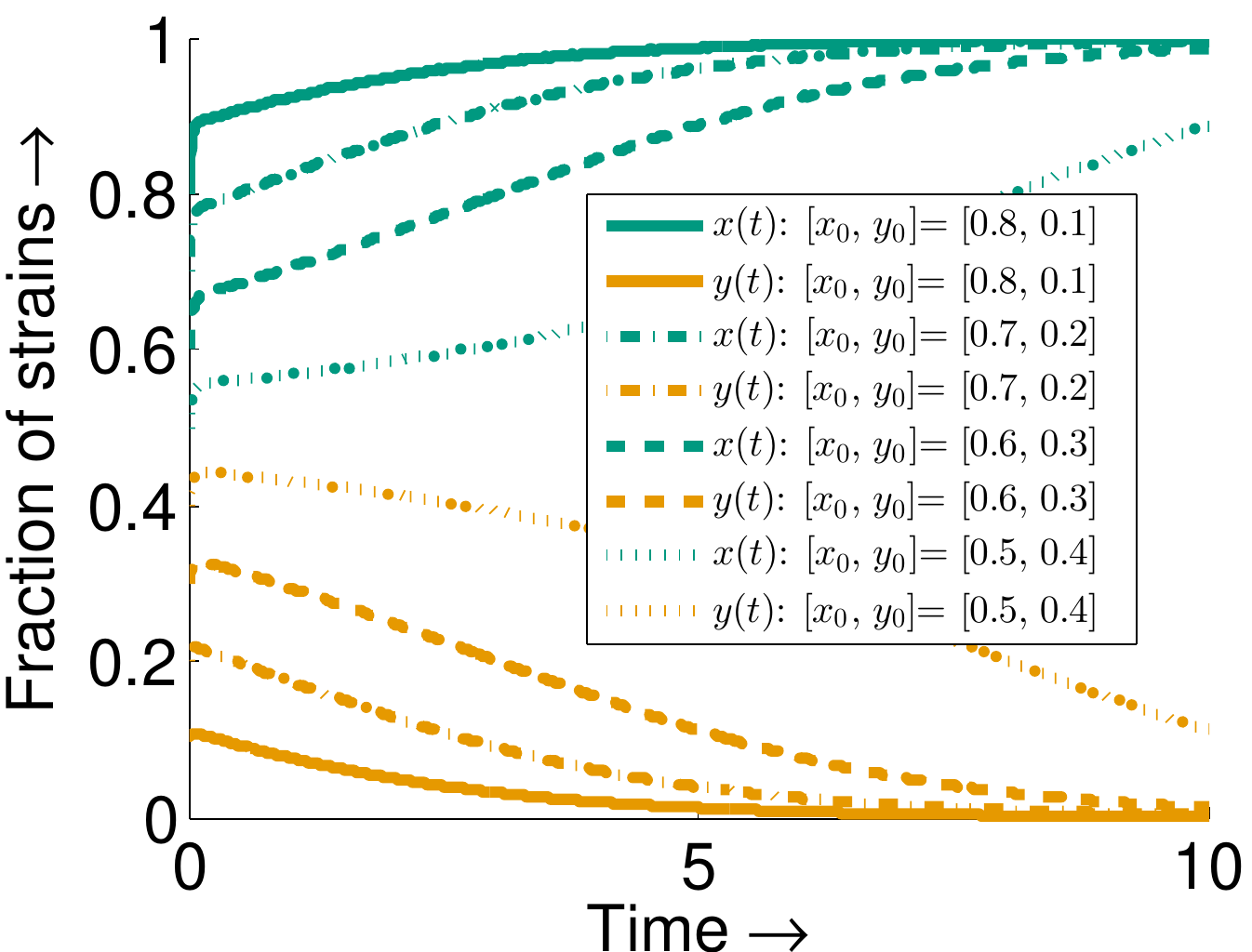} \\ 
\;\;\;\;\;\;\;\;\;\;\;\;\;\;\;\;\;\;\;\;\;\;\;\;c  & \;\;\;\;\;\;\;\;\;\;\;\;\;\;\;\;\;\;\;\;\;\;\;\;d 
\end{tabular}
\begin{tabular}{l}
\includegraphics[scale=0.345]{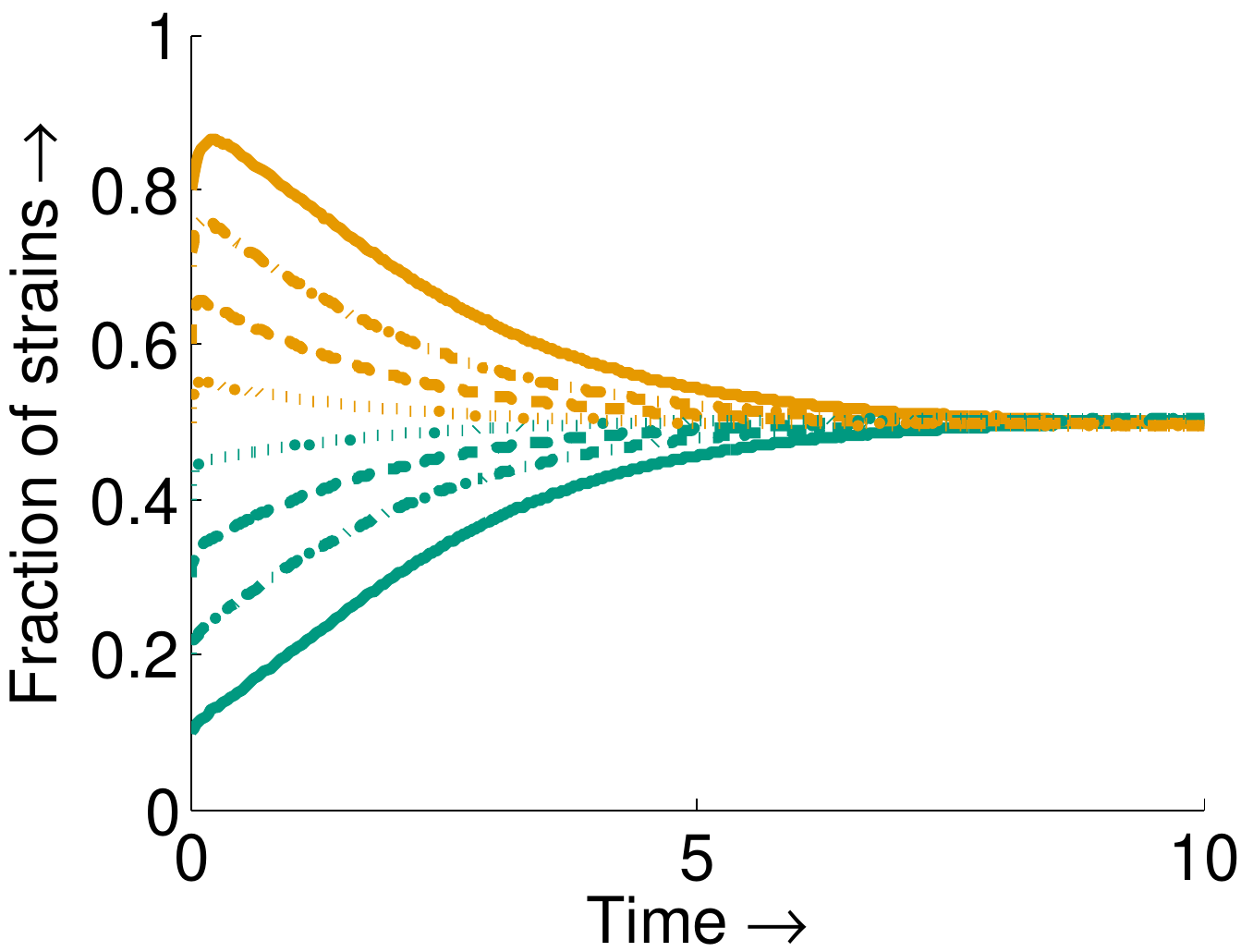}\\
\;\;\;\;\;\;\;\;\;\;\;\;\;\;\;\;\;\;\;\;\;\;\;\;\;\;\;\;\;e 
\end{tabular}
\caption{(Color online) Simulation of dynamics of $x$ and $y$ from (\ref{eq_dynamics_with_assortment}). (a) Defectors dominate for $r<N$, $\beta>\zeta$; (b) Cooperators dominate for $r>N$, $\zeta>\beta$; (c) and (d) The domination of cooperators or defectors depends on the relation between initial fraction and level of coexistence for $r<N$, $\zeta>\beta$; (e) Coexistence between cooperators and defectors for $r>N$, $\beta>\zeta$. Legend of figure (a) is applicable to the figures (b), (c) and (e).}
\label{Fig: dynamics_r_less_N_r_grt_N_with_zeta_beta_relation}
\end{figure}

\begin{figure}[here]
\centering
\begin{tabular}{ll}
\includegraphics[scale=0.28]{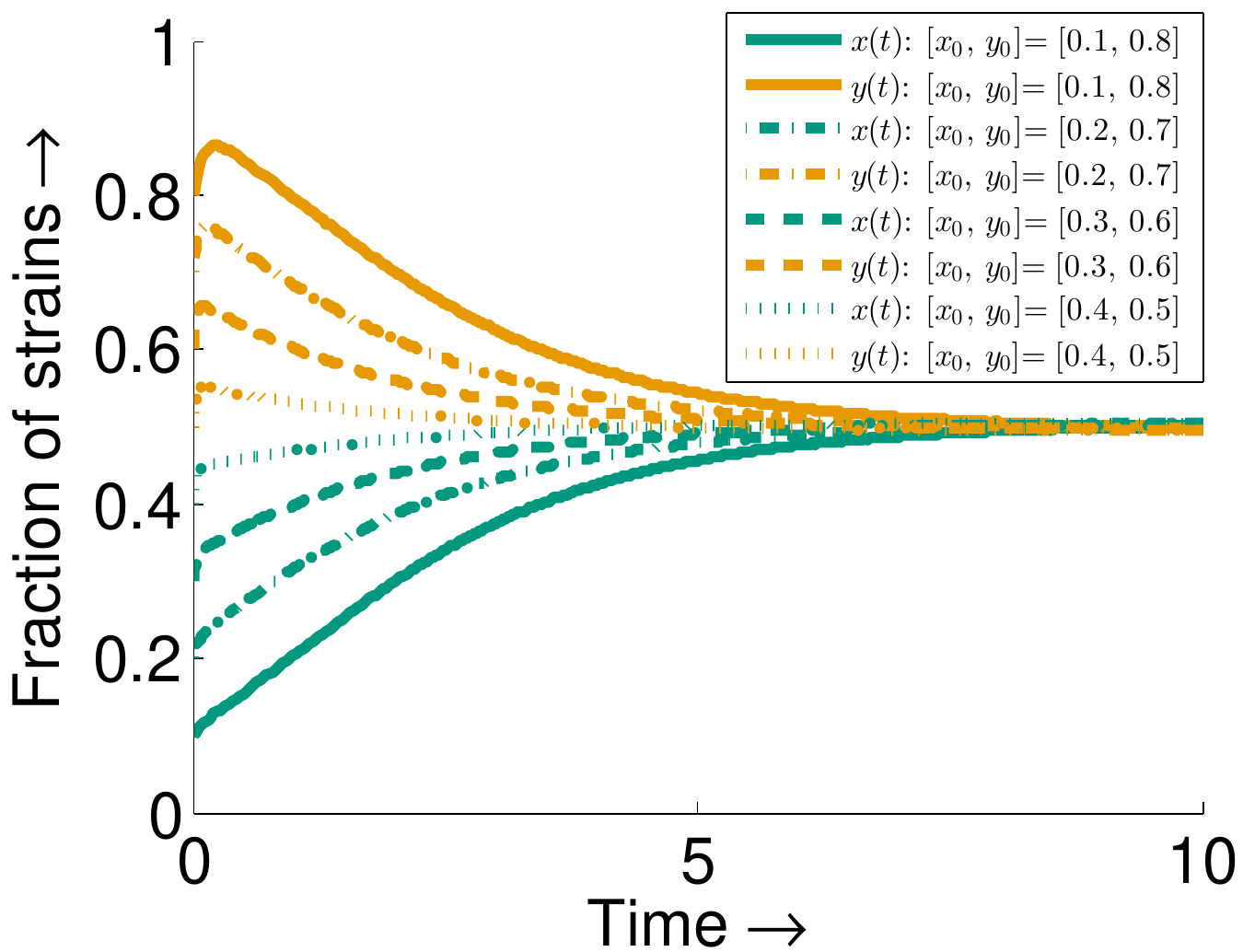} &  \includegraphics[scale=0.28]{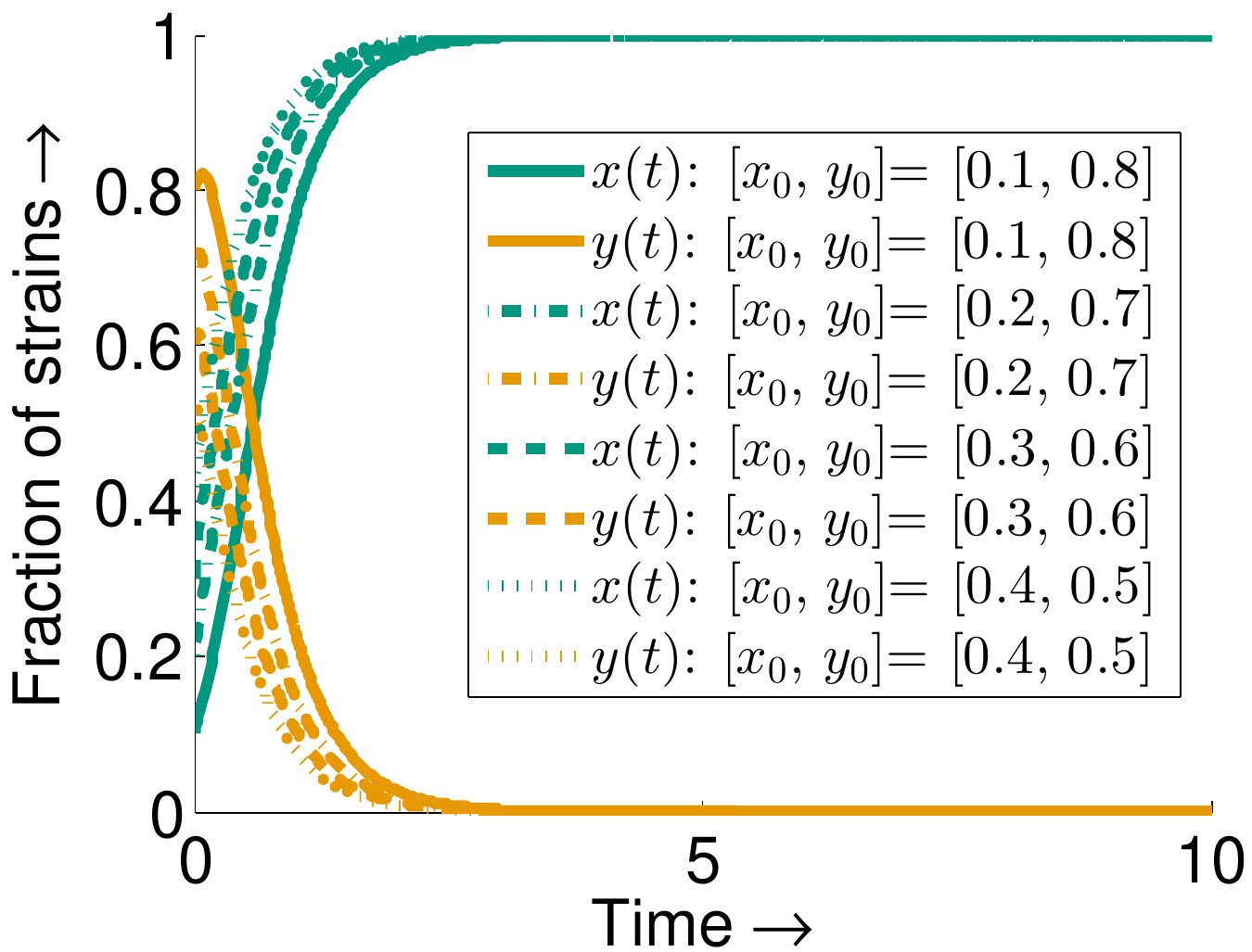}\\
\;\;\;\;\;\;\;\;\;\;\;\;\;\;\;\;\;\;\;\;\;\;\;\;a  & \;\;\;\;\;\;\;\;\;\;\;\;\;\;\;\;\;\;\;\;\;\;\;\;b\\
\end{tabular}
\caption{(Color online) Dynamics of $x$ and $y$ for the case IV. (a) Coexistence occurs with $(r,N,\zeta,\beta)=(100, 50, 0.05, 0.07)$ and (b) Cooperators dominate for $(r,N,\zeta,\beta)=(100, 20, 0.05, 0.07)$.}
\label{Fig: stability_check_coexistence}
\end{figure}

In the previous section, we discussed the stability analysis of the equilibrium points of the replicator system in (\ref{eqn: repli_x_y}) using the fitness functions as defined in (\ref{eqn: def_f_C_f_D}). In this stability analysis, the payoff expressions in (\ref{eqn: def_f_C_f_D}) do not contain any parameter(s) which signify the degree of assortment, present in microbial biofilms. As a result of the previous analysis, a single fixed point $(1,0)$ (`full dominance of cooperators') satisfies the condition for global asymptotic stability. In the present section, we extend the previous discussion via inclusion of degree of assortment in the fitness functions of cooperators and defectors. The purpose of including the spatial pattern in the proposed analysis is twofold: (a) emergence of cooperation in a population with phenotypic heterogeneity facilitates spatial patterns in microbial ecosystems \cite{ratzke2016self}, and (b) the degree of assortment can be controlled externally with the variation of initial cell density \cite{van2014density}. Hence, it could be interesting to visualize the effect of level of segregation on the population dynamics of cooperative and defective strains. Segregation in biofilms favors cooperative cell lines as the cooperators more often interact with their lineages, and high level of assortment resists exploitation of public goods by the noncooperative `cheaters' \cite{hibbing2010bacterial, nadell2010emergence}. On the contrary, defectors are advantaged in well-mixed colony. To define the fitness functions with the consideration of spatial patterns, as formulated in (\ref{eq_payoffs_with_assortment}), $\zeta$ and $\beta$ are added as an intrinsic advantage to the cooperators and defectors, respectively, and $\zeta,\beta>0$. Therefore, $\zeta>\beta$ defines segregated biofilms, whereas, $\beta>\zeta$ ensures well-mixed.        
\begin{align}
\begin{split}
f_{C}= & \bigg[\frac{rc}{N}(N-1)x-c\bigg(1-\frac{r}{N}\bigg)\bigg](1+\zeta)= (Ax-B)(1+\zeta)\\
f_{D}= & \bigg[\frac{rc}{N}(N-1)x\bigg](1+\beta)= Ax(1+\beta).
\end{split}
\label{eq_payoffs_with_assortment}
\end{align}
\indent By substituting the payoff expressions (\ref{eq_payoffs_with_assortment}) in the replicator equation (\ref{eqn: repli_x_y}), we can write the dynamics of cooperative and defective strains in a segregated or well-mixed colony with non-participating strains as 
\begin{align}
\begin{split}
\dot{x}= & x\big[(1-x)(Ax-B)(1+\zeta)-Axy(1+\beta)\big]\\
\dot{y} = & y\big[x\{A(1+\beta)-(Ax-B)(1+\zeta)\}-Axy(1+\beta)\big].
\end{split}
\label{eq_dynamics_with_assortment}
\end{align}
\indent The equilibrium points, $(x^{*},y^{*})$, are calculated from (\ref{eq_dynamics_with_assortment}) using $\dot{x}$=0 and $\dot{y}$=0 as $(0, 0)$, $\bigg(0, 1+\frac{B(1+\zeta)}{A(1+\beta)}\bigg)$, $(0, 1)$, $(1, 0)$, $\big(\frac{B}{A}, 0\big)$ and the point of coexistence $(x^{*}_{\text{c}}, y^{*}_{\text{c}})$= $\bigg(\frac{B(1+\zeta)}{A(\zeta-\beta)}, 1+\frac{B(1+\zeta)}{A(\beta-\zeta)}\bigg)$. The dynamics in (\ref{eq_dynamics_with_assortment}) contains infinite number of fixed points in the form $(0, y^{*})$ where $y^{*}\in [0,1]$, however, an ecological public goods game with non-participating strains (i.e. average fitness of the population is dominated by only the competing strains) does not hold any asymptotically stable equilibrium in the absence of public goods producers \cite{doebeli2004evolutionary}. Even though we analyze the fixed points $(0, 0)$, $\bigg(0, 1+\frac{B(1+\zeta)}{A(1+\beta)}\bigg)$ as these are directly calculated from the dynamics in (\ref{eq_dynamics_with_assortment}), and we consider $(0, 1)$ as it is a boundary value of the infinite set of equilibrium states. A notable difference between the dynamics in (\ref{eqn: repli_x_y_modi}) and (\ref{eq_dynamics_with_assortment}) is that the incorporation of the level of assortment induces a unique point of coexistence \cite{frey2010evolutionary, madeo2015game} between the competing strains.\\
\indent Four possible cases are considered for the analysis of population dynamics in (\ref{eq_dynamics_with_assortment}) - \textit{case I}: $r<N, \beta>\zeta$ implies return is less to the cooperators than the metabolic cost of cooperation in a well-mixed biofilm, \textit{case II}: $r>N, \zeta>\beta$ manifests return is greater than the cost of cooperation in a segregated biofilm, \textit{case III}: $r<N, \zeta>\beta$ denotes return is less than the meatbolic cost in an assorted biofilm, and \textit{case IV}: $r>N, \beta>\zeta$ signifies the return is greater than the cooperative cost in a well-mixed biofilm. Population dynamics of cooperators and defectors are simulated from (\ref{eq_dynamics_with_assortment}) for the case I to case IV in the scenario $x+y<1$, and corresponding results are depicted in Fig. \ref{Fig: dynamics_r_less_N_r_grt_N_with_zeta_beta_relation} for a set of parameters $r$, $N$, $\zeta$ and $\beta$. 
\\\indent From Fig. \ref{Fig: dynamics_r_less_N_r_grt_N_with_zeta_beta_relation} (a), it is clear that the lesser return to the cooperators than the metabolic cost of production in well-mixed biofilms (case I) favors defectors, whereas, the greater return than the cost of cooperation in segregated biofilms (case II) supports cooperators as depicted in (b). Fig. \ref{Fig: dynamics_r_less_N_r_grt_N_with_zeta_beta_relation} (c, d) present bistability (winning of cooperators or defectors depend on the initial fraction) when the return is less than the cost of cooperation in assorted biofilms (case III), and Fig. \ref{Fig: dynamics_r_less_N_r_grt_N_with_zeta_beta_relation} (e) illustrates coexistence in a well-mixed colony with the greater return to the cooperators compared to the metabolic cost of public goods production (case IV). However, $(0,1)$ is not an asymptotically stable equilibrium point for a replicator system with non-competing strains as discussed previously. Therefore, the stable $(0,1)$ as shown in Fig. \ref{Fig: dynamics_r_less_N_r_grt_N_with_zeta_beta_relation} (a) and (c) may not be stable for the other set of $(r,N,\zeta,\beta)$, even though, that set satisfies the relation for the cases I and III. The point of coexistence is also not a stable state in the entire domain of case IV as shown in Fig. \ref{Fig: stability_check_coexistence}. It is shown that coexistence persist for $(r,N,\zeta,\beta)=(100, 50, 0.05, 0.07)$, however, the dynamics moves to another steady state $(1,0)$ while $(r,N)=(100,20)$ and $(\zeta,\beta)$ is unchanged. Therefore, the stability conditions of the equilibrium points for the valid cases are necessary to investigate further. Conditions for the stability of the fixed states $(1,0)$ and $(x^{*}_{\text{c}}, y^{*}_{\text{c}})$ are discussed in the following subsection.\\
\subsection{Stability conditions of $(1,0)$ and $(x^{*}_{\text{c}}, y^{*}_{\text{c}})$}
\label{stability_condition}
The replicator dynamics in (\ref{eq_dynamics_with_assortment}) can be written alternatively as
\begin{align}
\begin{split}
\dot{x}= & -A(1+\zeta)x^{3}+[(A+B)(1+\zeta)-A(1+\beta)y]x^{2}-B(1+\zeta)x\\
\dot{y}= & -A(1+\zeta)x^{2}y+[A(1+\beta)+B(1+\zeta)]xy-A(1+\beta)xy^{2}.
\end{split}
\label{eq_dynamics_with_assortment_alter}
\end{align}
\indent In the next subsections, we investigate the stability conditions of the fixed points $(1,0)$ and $(x^{*}_{\text{c}}, y^{*}_{\text{c}})$ using linear stability analysis via calculating the Jacobian from (\ref{eq_dynamics_with_assortment_alter}).
\subsubsection{Stability condition for $(1,0)$}
\label{stability_condition_1_0}
The Jacobian matrix of $(1,0)$ is calculated from the dynamics in (\ref{eq_dynamics_with_assortment_alter}) as
\begin{equation}
J^{\prime}_{(1,0)}= \left[ \begin{array}{cc}
(B-A)(1+\zeta) & -A(1+\beta) \\
0 & (B-A)(1+\zeta) +A(1+\beta)
\end{array} \right].
\label{eqn: jaco_1_0_with_segre}
\end{equation}
\indent The eigen values of $J^{\prime}_{(1,0)}$ are $\lambda^{\prime (1,0)}_{1}= (B-A)(1+\zeta)= -(1+\zeta)c(r-1)$ and $\lambda^{\prime (1,0)}_{2}= (B-A)(1+\zeta) +A(1+\beta)=-(1+\zeta)c(r-1)+(1+\beta)(\frac{rc}{N}(N-1))$. $\lambda^{\prime (1,0)}_{1}$ is always $<0$ as $c,\zeta>0$ and $r>1$, whereas for the stability of $(1,0)$, $\lambda^{\prime (1,0)}_{2}<0$, which gives a unique inequality for the four individual cases as 
\begin{align}
\bigg(\frac{\zeta-\beta}{1+\zeta}\bigg)+\bigg(\frac{1+\beta}{1+\zeta}\bigg)\frac{1}{N}>\frac{1}{r}.
\label{eqn: stablity_cond_1_0_with_segre}
\end{align} 
\indent Hence, $(1,0)$ is a stable equilibrium if the inequality condition in (\ref{eqn: stablity_cond_1_0_with_segre}) holds, and $(1,0)$ is a saddle if $\bigg(\frac{\zeta-\beta}{1+\zeta}\bigg)+\bigg(\frac{1+\beta}{1+\zeta}\bigg)\frac{1}{N}<\frac{1}{r}$.
\subsubsection{Stability condition for ($x^{*}_{\text{c}}, y^{*}_{\text{c}}$)}
\label{stability_condition_xc_yc} 
\begin{figure}[htbp]
\centering
\begin{tabular}{lll}
\;\;\;\;\;\;\;\;\;\;t=0 & \;\;\;\;\;\;\;\;\;\;t=6 & \;\;\;\;\;\;\;\;\;t=12\\
\thickhline\noalign{\smallskip}
\includegraphics[scale=0.21]{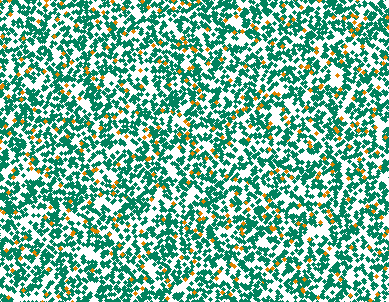} & \includegraphics[scale=0.21]{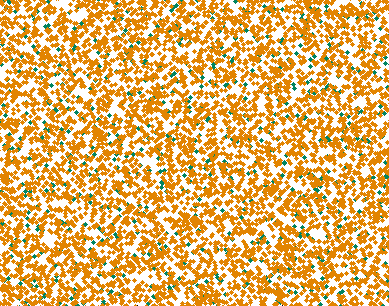} & \includegraphics[scale=0.21]{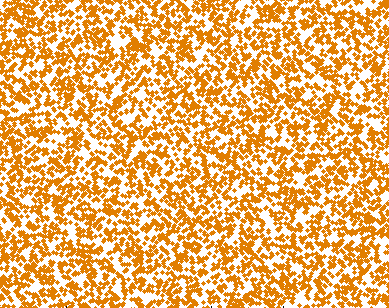} \\
\;\;\;\;\;\;\;\;\;\;\;a1  & \;\;\;\;\;\;\;\;\;\;\;a2  & \;\;\;\;\;\;\;\;\;\;\;a3  \\
\hline\noalign{\smallskip}
\includegraphics[scale=0.21]{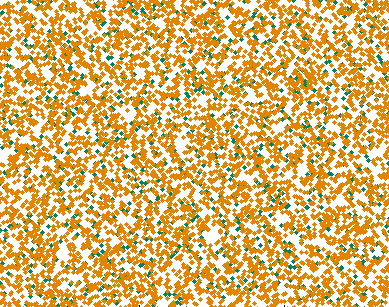} & \includegraphics[scale=0.21]{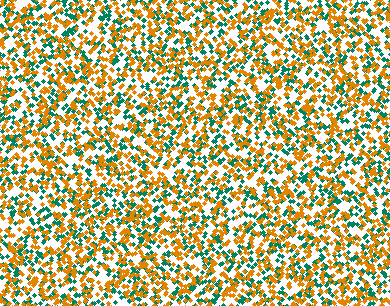} & \includegraphics[scale=0.21]{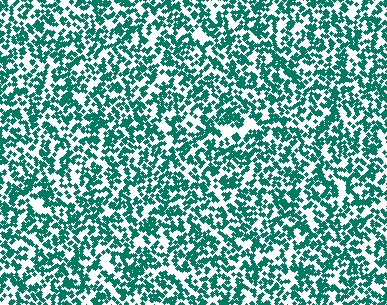} \\
\;\;\;\;\;\;\;\;\;\;\;b1  & \;\;\;\;\;\;\;\;\;\;\;b2  & \;\;\;\;\;\;\;\;\;\;\;b3 \\
\hline\noalign{\smallskip}
\includegraphics[scale=0.21]{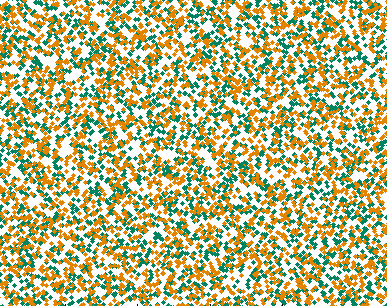} & \includegraphics[scale=0.21]{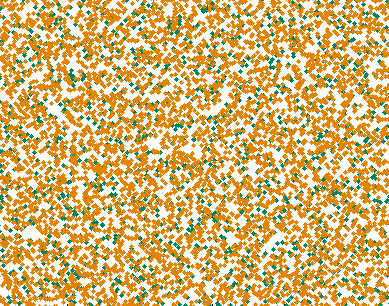} & \includegraphics[scale=0.21]{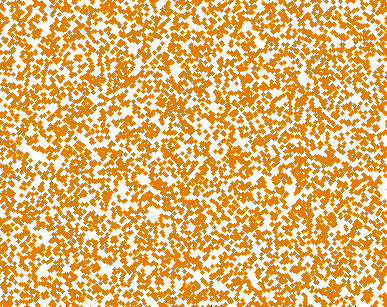} \\
\;\;\;\;\;\;\;\;\;\;\;c1  & \;\;\;\;\;\;\;\;\;\;\;c2  & \;\;\;\;\;\;\;\;\;\;\;c3 \\
\hline\noalign{\smallskip}
\includegraphics[scale=0.21]{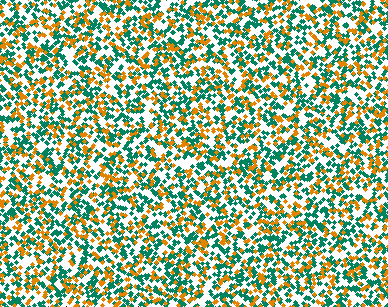} & \includegraphics[scale=0.21]{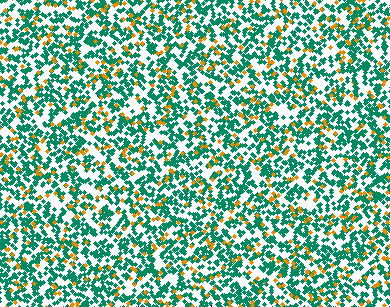} & \includegraphics[scale=0.21]{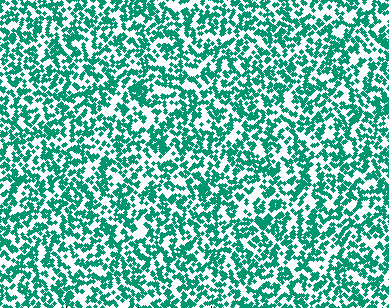} \\
\;\;\;\;\;\;\;\;\;\;\;d1  & \;\;\;\;\;\;\;\;\;\;\;d2  & \;\;\;\;\;\;\;\;\;\;\;d3 \\
\hline\noalign{\smallskip}
\includegraphics[scale=0.21]{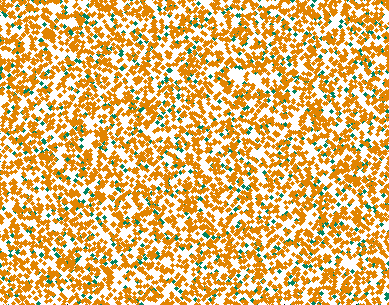} & \includegraphics[scale=0.21]{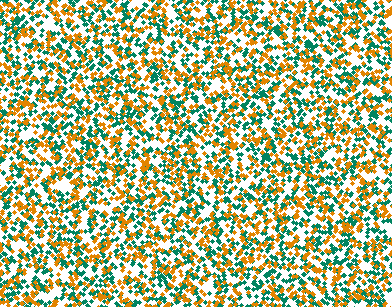} & \includegraphics[scale=0.21]{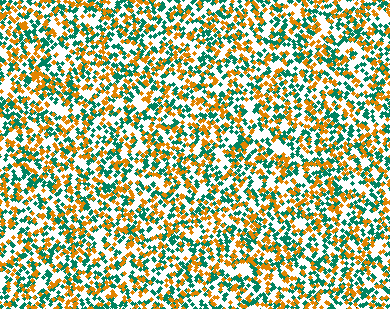} \\
\;\;\;\;\;\;\;\;\;\;\;e1  & \;\;\;\;\;\;\;\;\;\;\;e2  & \;\;\;\;\;\;\;\;\;\;\;e3\\ 
\thickhline\noalign{\smallskip}
\end{tabular}
\caption{(Color online) Snapshots from the video simulation of the four different cases for $x+y=1$. Time steps are mentioned at the top of the figure. (a1-a3) Case I: Defectors dominate, (b1-b3) Case II: Cooperators dominate, (c1-c3) Case III with initial fraction of cooperators as 0.4: Domination of defectors, (d1-d3) Case III with initial fraction of cooperators as 0.6: Domination of cooperators, (e1-e3) Case IV: Coexistence between cooperators and defectors. Initial fraction of cooperators is taken as 0.9, 0.1 and 0.1 for the case I, case II and case IV, respectively. \tikzcircle[fill=bluish_green_coop, bluish_green_coop]{4.1pt} and \tikzcircle[fill=orange_defec, orange_defec]{4.1pt} denote cooperator and defector, respectively.}
\label{Fig: dynamics_x_y_1_with_segre}
\end{figure}

The point of coexistence ($x^{*}_{\text{c}}, y^{*}_{\text{c}}$) is an invalid fixed point for the case I and case II. The Jacobian for $(x^{*}_{\text{c}}, y^{*}_{\text{c}})$ is calculated from the dynamics in (\ref{eq_dynamics_with_assortment_alter}) as
\begin{equation}
J^{\prime}_{(x^{*}_{\text{c}}, y^{*}_{\text{c}})}= \left[ \begin{array}{cc}
 -C_{1}+C_{4}&  -C_{2}\\\\
C_{1}+C_{3}& C_{2}\bigg(1+\frac{C_{3}}{C_{1}}\bigg)
\end{array} \right],
\label{eqn: jaco_xc_yc_with_segre}
\end{equation}
where $C_{1}=\frac{B^2(1+\zeta)^3}{A(\beta-\zeta)^2}$, $C_{2}=C_{1}\big(\frac{1+\beta}{1+\zeta}\big)$, $C_{3}=\frac{B(1+\zeta)^2}{\beta-\zeta}$, $C_{4}=B(1+\zeta)$.\\
\indent The eigen values of $J^{\prime}_{(x^{*}_{\text{c}}, y^{*}_{\text{c}})}$ are $\lambda^{\prime (x^{*}_{\text{c}}, y^{*}_{\text{c}})}_{1}=\frac{(1+\beta)(1+\zeta)B}{\beta-\zeta}$ and $\lambda^{\prime (x^{*}_{\text{c}}, y^{*}_{\text{c}})}_{2}=\frac{B^2(1+\zeta)^2}{A(\beta-\zeta)}+B(1+\zeta)$. $\lambda^{\prime (x^{*}_{\text{c}}, y^{*}_{\text{c}})}_{1}<0$ for the cases III and IV. For case III, $\lambda^{\prime (x^{*}_{\text{c}}, y^{*}_{\text{c}})}_{2}<0$ and $x^{*}_{\text{c}}< 1$ gives the inequality $\frac{N-r}{\zeta-\beta}>\frac{r(N-1)}{1+\zeta}$ and $\frac{N-r}{\zeta-\beta}<\frac{r(N-1)}{1+\zeta}$, respectively which are contradicting each other. Hence, for case III, $\lambda^{\prime (x^{*}_{\text{c}}, y^{*}_{\text{c}})}_{2}>0$, which intends that the point of coexistence is a saddle-node for $r<N, \zeta>\beta$. However, for case IV, $\lambda^{\prime (x^{*}_{\text{c}}, y^{*}_{\text{c}})}_{2}<0$ and $x^{*}_{\text{c}}< 1$ denote the same inequality condition as 
\begin{align}
\frac{N-r}{\zeta-\beta}<\frac{r(N-1)}{1+\zeta}.
\label{eqn: stablity_cond_xc_yc_with_segre}
\end{align}
\indent Therefore, $(x^{*}_{\text{c}}, y^{*}_{\text{c}})$ is a stable steady state if the inequality condition in (\ref{eqn: stablity_cond_xc_yc_with_segre}) satisfies for the case IV.\\
\indent A system with only cooperative and defective strains is a special case of our previous discussion with non-participating strains. For $x+y=1$, full dominance of cooperators and defectors ($(1,0)$ and $(0,1)$, respectively) are asymptotically stable for the case I and case II, respectively. Both $(1,0)$ and $(0,1)$ are stable for case III (bistability) if the inequality $\frac{\zeta-\beta}{(1+\zeta)}-\frac{(N-r)}{r(N-1)}>0$ holds, and the point of coexistence $(x^{*}_{\text{c}}, 1-x^{*}_{\text{c}})$ is stable for the case IV upon satisfying the inequality condition $\frac{\zeta-\beta}{(1+\zeta)}-\frac{(N-r)}{r(N-1)}<0$. The population dynamics of cooperators for $x+y=1$ is simulated from (\ref{eq_dynamics_with_assortment}) by substituting $y=1-x$, and corresponding snapshots from the video simulation are depicted in Fig. \ref{Fig: dynamics_x_y_1_with_segre}. The results in Fig. \ref{Fig: dynamics_x_y_1_with_segre} satisfy the discussed stable steady states for the four cases in a special scenario $x+y=1$.\\
\indent The summary of stability analysis of the fixed points for the system in (\ref{eq_dynamics_with_assortment}) $(x+y<1)$ is tabulated in Table \ref{tab_summary_with_segre}. The stability analysis of the equilibrium states other than $(1,0)$ and ($x^{*}_{\text{c}}, y^{*}_{\text{c}}$) are discussed in the supplementary material.
\begin{table*}[t]
\begin{center}
\caption{Summary of stability analysis of fixed points with assortment}
\begin{tabular}{l"llll}
\thickhline\noalign{}
\;\;\;\;\;\;\;\textbf{Fixed points} & \multicolumn{4}{c}{\textbf{Nature of fixed points}}\\
\thickhline\noalign{}
& \begin{tabular}{@{}c@{}}Case I\\$r<N, \beta>\zeta$\end{tabular} & \begin{tabular}{@{}c@{}}Case II\\$r>N, \zeta>\beta$\end{tabular} & \begin{tabular}{@{}c@{}}Case III\\$r<N, \zeta>\beta$\end{tabular}  & \begin{tabular}{@{}c@{}}Case IV\\$r>N, \beta>\zeta$\end{tabular}\\
\thickhline\noalign{}
\;\;\;\;\;\;\;\;\;\;\;\;\;(0,0) & Unstable node & Unstable node & Unstable node & Unstable node\\
\hline\noalign{}
\;\;\;\;\;$\bigg(0,1+\frac{B(1+\zeta)}{A(1+\beta)}\bigg)$ & Not valid & Unstable node & Not valid & Unstable node\\
\hline\noalign{}
\;\;\;\;\;\;\;\;\;\;\;\;\;\textbf{(1,0)} & \multicolumn{4}{c}{\begin{tabular}{@{}c@{}}\textbf{Stable node}\\\textbf{if (\ref{eqn: stablity_cond_1_0_with_segre}) holds true}\end{tabular}}\\
\hline\noalign{}
\;\;\;\;\;\;\;\;\;\;\;\;$(\frac{B}{A},0)$ & Unstable node & Not valid & Unstable node & Not valid\\
\hline\noalign{}
\;\;\;\;\;\;\;\;\;\;\;\;\;(0,1) & Unstable node & Unstable node & Unstable node & Unstable node\\
\hline\noalign{}
$\bigg(\frac{\bm {B(1+\zeta)}}{\bm {A(\zeta-\beta)}}, \bm {1+\frac{\bm {B(1+\zeta)}}{\bm {A(\beta-\zeta)}}}\bigg)$ & Not valid & Not valid & \textbf{Saddle-node} &\begin{tabular}{@{}c@{}}\textbf{Stable node}\\\textbf{if (\ref{eqn: stablity_cond_xc_yc_with_segre}) holds true}\end{tabular}\\
\thickhline\noalign{}
\end{tabular}
\label{tab_summary_with_segre} 
\end{center}
\end{table*}
\subsection{Inference from the above analysis}
\label{infer_dynamics_with_segre}
\begin{figure}[htbp]
\centering
\includegraphics[scale=0.55]{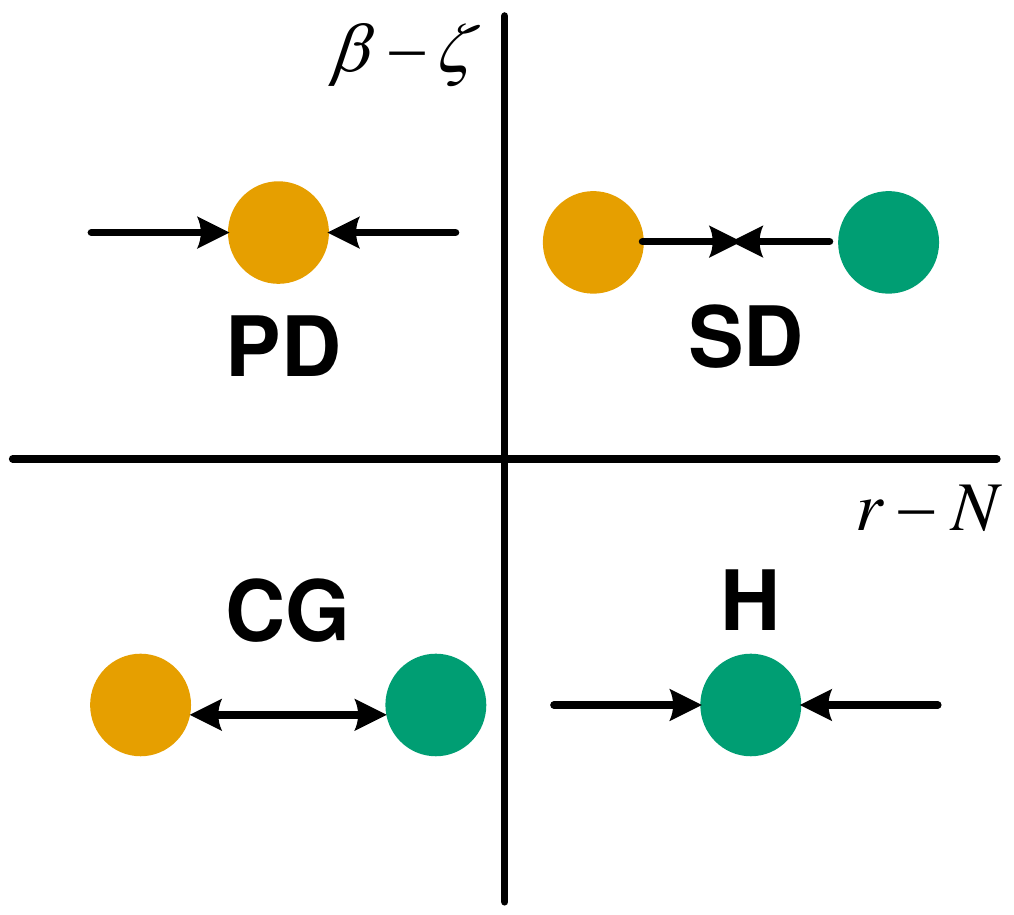} 
\caption{(Color online) Schematic diagram of four possible games in biofilms. Case I ($r<N,\beta>\zeta$) defines the winning of defectors, can be modeled as PD game. Case II ($r>N,\zeta>\beta$) signifies the domination of cooperators which resembles H game. Case III ($r<N,\zeta>\beta$) shows bistability, similar to CG. And case IV ($r>N,\beta>\zeta$) denotes coexistence as similar to SD game.  \tikzcircle[fill=bluish_green_coop, bluish_green_coop]{4.1pt} and \tikzcircle[fill=orange_defec, orange_defec]{4.1pt} denote the subpopulation of cooperator and defector, respectively.}
\label{Fig: four_games}
\end{figure}

\indent $\bullet${ Consideration of the degree of assortment \cite{van2014density} introduces bistability \cite{frey2010evolutionary} and coexistence \cite{frey2010evolutionary} in a microbial biofilm. When return to the cooperators is less than the cost of production in a segregated colony, the dominance of cooperators or defectors depends on its initial population density (case III: bistability). Whereas the colony goes into a stable mixed steady-state (case IV: coexistence) when the return is more in a well-mixed biofilm. Coexistence occurs as the perfect defection (i.e. cooperators stop the production of public goods or defectors fully exploit the goods produced by the cooperators) is not a suitable strategy towards the long-term existence for both the strains \cite{doebeli2004evolutionary}.}\\
\indent $\bullet${ The replicator system \cite{frey2010evolutionary} with the inclusion of both spatial pattern \cite{van2014density} and non-competing strains \cite{czaran2009microbial} does not show global asymptotic stability. As a result, we got the conditions for the dominance of cooperators and coexistence. Therefore, there is a scope to regulate the system parameters (here $r$, $N$, $c$, $\zeta$, $\beta$) such that the inequality condition (equation (\ref{eqn: stablity_cond_1_0_with_segre})) for the appropriation of pathogenic (from the viewpoint of hosts) cooperators violates - can be a novel strategy for the eradication of infectious biofilms.} \\ 
\indent $\bullet${ The outcomes of four cases (see Fig. \ref{Fig: dynamics_r_less_N_r_grt_N_with_zeta_beta_relation} and Fig. \ref{Fig: dynamics_x_y_1_with_segre}) resemble four different games \cite{frey2010evolutionary} - prisoners' dilemma (PD), harmony (H), coordination game (CG) and snowdrift game (SD) as shown in Fig. \ref{Fig: four_games}.}
\section{Conclusion}
\label{conclu}
Phenotypic and genotypic heterogeneity in the bacterial colony is a driving force for analyzing the dynamics of the constituent population \cite{wei2015molecular}. Therefore, the impact of social interactions \cite{oliveira2015biofilm, kim2016rapid, yurtsev2016oscillatory} between these heterotypic strains is obligatory for the investigation of microbial population dynamics \cite{wei2015molecular, li2015games}. In addition, stability analysis of the population structure is one of the best ways to find the underlying dominant strategy present in the colony \cite{li2015games}. In the support of the above statements, in this paper, we analyze the inherent population dynamics of a microbial biofilm in a combined framework of EGT \cite{frey2010evolutionary} and nonlinear control theory \cite{strogatz2014nonlinear}. Our analysis of this replicator system with ``inactive alleles'' \cite{czaran2009microbial} is twofold: with consideration of spatial pattern \cite{van2014density} and without inclusion of the assortment level. Former shows the `full dominance of pathogenic cooperators' as a global asymptotic stable equilibrium which justifies the adversity in the eradication of infectious biofilms. Whereas the latter introduces two additional cases - bistability and coexistence, and also presents the conditional dominance of cooperation and coexistence. Here, the outcomes of four possible cases resemble four different evolutionary games \cite{frey2010evolutionary}. The stability analysis developed for the proposed sociomicrobiological system not only presents a standard framework for analyzing multi-order, nonlinear population dynamics but also gives a clue to move the system dynamics to a desired steady state with external supervision (drug) for controlling the bacterial pathogenicity \cite{bjarnsholt2013role, xavier2016sociomicrobiology}.\\
\indent In future, we will work on the modeling and analysis of the replicator system for an inhomogeneous and variable sized population. Further investigation is required to find out the significance of $r>1$ and $r>N$ in biofilms, and for this analysis, we may need to include non-competing-type public goods producers (e.g. altruists) in the microbial population. One major challenge is to verify this analysis through wet lab experiments by tagging cooperators and defectors with fluorescent proteins. This would help in observing the dynamics, and provide valuable insights for further analysis of the system as well as assist in designing better therapeutic control measures.  
\section{Acknowledgments}
The work has been supported by the project `WBC' funded by Indian Institute of Technology Kharagpur, India. The authors would like to extend their gratitude to the editor and the anonymous referees for taking their invaluable time to improve this paper. 

\bibliographystyle{abbrv}
\bibliography{sigproc}  
%
%
\appendix
\section{Supplementary material}
\label{supplemen}
\subsection{Stability analysis of equilibrium points with spatial pattern}
\label{stab_analy_remaining_with_segre}

\subsubsection{Analysis of $(0,0)$ and $(0,1)$}
\label{analy_0_0_with_segre}
The Jacobian matrix for the fixed point $(0,0)$ is calculated from the dynamics in (\ref{eq_dynamics_with_assortment_alter}) as
$
J^{\prime}_{(0,0)}= \left[ \begin{array}{cc}
-B(1+\zeta) & 0 \\
0 & 0
\end{array} \right].
$
The eigen values of $J^{\prime}_{(0,0)}$ are $\lambda^{\prime (0,0)}_{1}=-B(1+\zeta)=-c(1-\frac{r}{N})(1+\zeta)$ and $\lambda^{\prime (0,0)}_{2}=0$. The eigen vectors corresponding to $\lambda^{\prime (0,0)}_{1}$ and $\lambda^{\prime (0,0)}_{2}$ are $y=0$ and $x=0$, respectively. For $r<N$, $\lambda^{\prime (0,0)}_{1}<0$ and for $r>N$, $\lambda^{\prime (0,0)}_{1}>0$.    
To say about the nature of eigen vector $x=0$ for $\lambda^{\prime (0,0)}_{2}=0$, a parabolic manifold centre at $(0,0)$ is taken as $x=h(y)=h_{2}y^{2}$. The value of $h_{2}$ is calculated from (\ref{eqn: fr_calc_h_2}) by substituting $x=h_{2}y^{2}$. We get the parameters $A^{\prime}$, $B^{\prime}$, $f^{\prime}$ and $g^{\prime}$ in (\ref{eqn: fr_calc_h_2}) by comparing (\ref{eqn: dyna_cen_mani}) and (\ref{eq_dynamics_with_assortment_alter}) as $A^{\prime}=-B(1+\zeta)$, $B^{\prime}=0$, $f^{\prime}=-A(1+\zeta)x^{3}+[(A+B)(1+\zeta)-A(1+\beta)y]x^{2}$ and $g^{\prime}= -A(1+\zeta)x^{2}y+[A(1+\beta)+B(1+\zeta)]xy-A(1+\beta)xy^{2}$. Hence, the expression from (\ref{eqn: fr_calc_h_2}) can be written as
\begin{align}
\begin{split}
(1+\zeta)Bh_{2}y^{2}+\mathcal{O}(y^{4},y^{5},y^{6})=0
\end{split}
\label{eq: expre_h2_0_0}
\end{align}   
\indent Similarly, we get the Jacobian from (\ref{eq_dynamics_with_assortment_alter}) for $(0,1)$ as\\
$
J^{\prime}_{(0,1)}= \left[ \begin{array}{cc}
-B(1+\zeta) & 0 \\
B(1+\zeta) & 0
\end{array} \right].
$
The eigen values of $J^{\prime}_{(0,1)}$ are $\lambda^{\prime (0,1)}_{1}=-B(1+\zeta)=-c(1-\frac{r}{N})(1+\zeta)$ and $\lambda^{\prime (0,1)}_{2}=0$, and corresponding eigen vectors are $y=-x$ and $x=0$, respectively. The eigen vector $y=-x$ converges to $(0,1)$ for $r<N$ and diverges for $r>N$. As similar to the analysis of the fixed point $(0,0)$, a parbolic manifold centre at $(0,1)$, $x=h(y)=h_{2}(y-1)^2$, is taken to denote the nature of eigen vector $x=0$ for $\lambda^{\prime (0,1)}_{2}=0$. Hence, after neglecting h.o.ts $(>\mathcal{O}(y^{2}))$, the expression from (\ref{eqn: fr_calc_h_2}) can be written as
\begin{align}
\begin{split}
& y^{2}[5A(1+\zeta)h_{2}^{3}+\{8A(1+\beta)+(4B-2A)(1+\zeta)\}h_{2}^{2}+B(1+\zeta)h_{2}]\\& +y[-4A(1+\zeta) h_{2}^{3}-\{A(1+\beta)+2B(1+\zeta)\}h_{2}^{2}-2(1+\zeta)Bh_{2}]+\\
& [A(1+\zeta)h_{2}^{3}-(A+B)(1+\zeta)h_{2}^{2}+B(1+\zeta)h_{2}]=0
\end{split}
\label{eq: expre_h2_0_1}
\end{align} 
The expressions in (\ref{eq: expre_h2_0_0}) and (\ref{eq: expre_h2_0_1}) are equal to zero if $h_{2}=0$, therefore, the centre manifold for both the equilibrium points $(0,0)$ and $(0,1)$ is $x=0$, which intends the dynamics $\dot{x}=0$ and $\dot{y}=0$ as calculated from (\ref{eq_dynamics_with_assortment_alter}). Hence, due to the constant nature of population density, it can be stated that fixed points $(0,0)$ and $(0,1)$ are unstable for any parameters set of $(r,N,\zeta,\beta)$ in a colony with non-participating strains.     
\subsubsection{Analysis of $\bigg(0,1+\frac{B(1+\zeta)}{A(1+\beta)}\bigg)$ and $(\frac{B}{A},0)$}
\label{analy_others_with_segre}
The equilibrium point $\bigg(0,1+\frac{B(1+\zeta)}{A(1+\beta)}\bigg)$ is represented here as $(0,1+p)$, where $p=\frac{B(1+\zeta)}{A(1+\beta)}$. The condition to make $(0,1+p)$ as a valid fixed point is $p=\frac{(1+\zeta)(N-r)}{(1+\beta)r(N-1)}<0$. Hence, $(0,1+p)$ is an invalid fixed point for the case I and case III. The Jacobian for $(0,1+p)$ can be denoted as
$
J^{\prime}_{(0,1+p)}= \left[ \begin{array}{cc}
-B(1+\zeta) & 0 \\
0 & 0
\end{array} \right].
$
$J^{\prime}_{(0,1+p)}$ is same as $J^{\prime}_{(0,0)}$. Hence, further analysis of $(0,1+p)$ using centre manifold theory will be the same as discussed in \ref{analy_0_0_with_segre}. The only difference - manifold expression for $(0,1+p)$ is $h_{2}(y-1-p)^2$. The further analysis defines $(0,1+p)$ as an unstable fixed point.\\
\indent The equilibrium point $(\frac{B}{A},0)$ is a valid steady state for case I and case III as $\frac{B}{A}=\frac{N-r}{r(N-1)}<0$ for the cases II and IV. The inequality $\frac{B}{A}< 1$ denotes $N(r-1)> 0$, which is always true for $r,N>1$, hence, for the cases I and III, $0<\frac{B}{A}< 1$. The Jacobian for $(\frac{B}{A},0)$ is expressed from the dynamics (\ref{eq_dynamics_with_assortment_alter}) as
$
J^{\prime}_{(\frac{B}{A},0)}= \left[ \begin{array}{cc}
B(1-\frac{B}{A})(1+\zeta) & -\frac{B^{2}}{A}(1+\beta) \\
0 & B(1+\beta)
\end{array} \right].
$     
The eigen values of $J^{\prime}_{(\frac{B}{A},0)}$ are $\lambda^{\prime (\frac{B}{A},0)}_{1}=\frac{(1+\zeta)c(r-1)(N-r)}{r(N-1)}$ and $\lambda^{\prime (\frac{B}{A},0)}_{2}=(1+\beta)c(1-\frac{r}{N})$. For $r<N$, $\lambda^{\prime (\frac{B}{A},0)}_{1}>0$ and $\lambda^{\prime (\frac{B}{A},0)}_{2}>0$, therefore, it can be concluded that $(\frac{B}{A},0)$  is an unstable fixed point for case I and case III. 
\end{document}